\documentclass[preprint,12pt,authoryear]{elsarticle}

\usepackage{amssymb}
\usepackage{amsmath}
\usepackage{url}
\usepackage{xcolor}
\usepackage{siunitx}
\usepackage{hyperref}
\usepackage{amsfonts,hyperref,bm,subcaption,multirow}
\usepackage{algorithmic}
\usepackage{graphicx}
\usepackage{textcomp}
\usepackage{float}
\usepackage{lineno}

\usepackage[algoruled,vlined,linesnumbered]{algorithm2e}

\SetKwInput{KwInput}{Input}                % Set the Input
\SetKwInput{KwOutput}{Output}              % set the Output

\begin{document}

\begin{frontmatter}



\title{CycleULM: A unified label-free deep learning framework for ultrasound localisation microscopy}

\author[1]{Su Yan}
\author[1]{Clara Rodrigo Gonzalez}
\author[2]{Vincent C. H. Leung}
\author[3]{Herman Verinaz-Jadan}
\author[2]{Jiakang Chen}
\author[1]{Matthieu Toulemonde}
\author[1]{Kai Riemer}
\author[1]{Jipeng Yan}
\author[1]{Clotilde Vi\'e}
\author[1]{Qingyuan Tan}
\author[1]{Peter D. Weinberg}
\author[2]{Pier Luigi Dragotti}
\author[4]{Kevin G. Murphy}
\author[1]{Meng-Xing Tang\corref{cor1}}
\cortext[cor1]{Corresponding author. Tel.: +44-2075943664}
\ead{mengxing.tang@imperial.ac.uk}

\address[1]{Department of Bioengineering, Faculty of Engineering, Imperial College London, London, UK.}
\address[2]{Department of Electrical and Electronic Engineering, Faculty of Engineering, Imperial College London, London, UK.}
\address[3]{Faculty of Electrical and Computer Engineering, Escuela Superior Polit\'ecnica del Litoral (ESPOL), Guayaquil, Ecuador.}
\address[4]{Department of Metabolism, Digestion and Reproduction, Faculty of Medicine, Imperial College London, London, UK.}

%% Abstract
\begin{abstract}

Super-resolution ultrasound via microbubble (MB) localisation and tracking, also known as ultrasound localisation microscopy (ULM), can resolve microvasculature beyond the acoustic diffraction limit. However, significant challenges remain in localisation performance and data acquisition and processing time. Deep learning methods for ULM have shown promise to address these challenges, however, they remain limited by in vivo label scarcity and the simulation-to-reality domain gap. We present CycleULM, the first unified label-free deep learning framework for ULM. CycleULM learns a physics-emulating translation between the real contrast-enhanced ultrasound (CEUS) data domain and a simplified MB-only domain, leveraging the power of CycleGAN without requiring paired ground truth data. With this translation, CycleULM removes dependence on high-fidelity simulators or labelled data, and makes MB localisation and tracking substantially easier. Deployed as modular plug-and-play components within existing pipelines or as an end-to-end processing framework, CycleULM delivers substantial performance gains across both in silico and in vivo datasets. Specifically, CycleULM improves image contrast (contrast-to-noise ratio) by up to 15.3 dB and sharpens CEUS resolution with a 2.5× reduction in the full width at half maximum of the point spread function. CycleULM also improves MB localisation performance by up to +40\% recall, +46\% precision, and $-$14.0 µm mean localisation error, yielding more faithful vascular reconstructions. Importantly, CycleULM achieves real-time processing throughput at 18.3 frames per second with order-of-magnitude speed-ups (up to $\sim$14.5×). By combining label-free learning, performance enhancement, and computational efficiency, CycleULM provides a practical pathway toward robust, real-time ULM and accelerates its translation to clinical applications.

\end{abstract}

\end{frontmatter}


Single-molecule localisation microscopy (SMLM) is a super-resolution optical imaging technique that breaks the diffraction limit by localising the stochastic blinking of fluorophore emissions, achieving order-of-magnitude gains in spatial resolution \citep{betzig2006imaging,rust2006sub}. Translating this localisation principle to ultrasound has led to super-resolution ultrasound (SRUS), also known as ultrasound localisation microscopy (ULM). In ULM, intravenously injected microbubble (MB) contrast agents are localised and tracked over time, enabling in vivo microvascular imaging with spatial resolution beyond the acoustic diffraction limit \citep{christensen2014vivo,errico2015ultrafast}. ULM has now been demonstrated across a wide range of settings, from in vitro phantoms \citep{couture2011microbubble,viessmann2013acoustic,desailly2013sono} to preclinical in vivo applications in tumour models, brain, kidney, heart, spinal cord and lymphatic vasculature in rodents and rabbits \citep{siepmann2011imaging,lin20173,christensen2014vivo,errico2015ultrafast,couture2018ultrasound,renaudin2022functional,sui2022randomized,guo2023frame,bar2018sushi,song2017improved,zhu20193d,yu2023super,demeulenaere2022coronary,taghavi2022ultrasound,lin2024super,lowerison2024super}. More recently, human studies have reported ULM of the breast, lower limb, liver, brain, lymph nodes, heart, kidney and carotid artery \citep{opacic2018motion,porte2024ultrasound,harput2018two,huang2021super,zeng2024focal,demene2021transcranial,denis2025transcranial,zhu2022super,yan2024transthoracic,huang2025multiphase,mandelbaum2025sensing,leroy2025assessment}. ULM has also been extended to three-dimensional (3D) imaging, further expanding its clinical potential \citep{foroozan2018microbubble,heiles2019ultrafast,demeulenaere2022vivo,wu20243d,hansen2024ultrafast,bureau2025ultrasound}, and to non-contrast enhanced ultrasound \citep{jensen2024super,naji2024super}.

Despite these recent advances, ULM still faces significant methodological challenges: (i) separating MB signals from strong background noise remains difficult, (ii) overlapping MBs degrade localisation, (iii) long acquisition times increase motion artefacts, and (iv) large data volumes require heavy computational loads and long processing times. Although deep learning (DL) shows promise for ULM by learning complex spatiotemporal representations \citep{rauby2024deep}, a major barrier for such DL appraoches is the lack of labelled in vivo data.  Manual annotation of MB locations and trajectories in in-vivo contrast-enhanced ultrasound (CEUS) frames is infeasible at scale, and pseudo-labels derived from conventional algorithms inevitably inherit their errors and biases. To circumvent this, most networks are trained on synthetic MB data generated with various ultrasound simulators, including Gaussian point spread function (PSF) models \citep{van2020super,liu2020deep,liu2023ultrasound,luan2023deep,zhang2024vivo}, Field II \citep{chen2022deep,youn2020detection,stevens2022hybrid,chen2023localization}, and SIMUS \citep{milecki2021deep}. However, these simulators cannot fully replicate real-world conditions, including acoustic wave propagation, nonlinear MB responses, the imaging system, and heterogeneous tissue properties \citep{pinton2011sources,versluis2020ultrasound,shin2024context}, resulting in a domain gap between the simulated data and the real in vivo data. This gap degrades the model performance when models trained on simulations are applied to real acquisitions \citep{rauby2024deep}, thereby hindering their widespread in vivo and clinical applications. To narrow the gap, a recent study trained a generative adversarial network (GAN) to generate more realistic MB signals, using the isolated MB patches manually selected from the in vivo dataset \citep{shin2024context}. However, stable GAN training requires a large data set, making manual data preparation labour-intensive. In addition, interactions between overlapping MBs are not considered in the model, limiting the realism of the simulated signals. 

In this study, we present CycleULM, the first unified, label-free DL framework for ULM post-processing. CycleULM eliminates dependence on high-fidelity simulators or labelled real data by learning, in a self-supervised manner, a bidirectional, physics-emulating domain translation from CEUS frames to a simplified MB-only domain via cycle-consistent adversarial training. This translation closes the simulation-to-reality gap and enables downstream localisation and tracking to be trained and operated in a controlled, interpretable domain with known PSFs and trajectories. Moreover, the CycleULM framework adopts a modular design comprising three neural networks: the Microbubble Domain Translator (MB-DT) for MB isolation and speckle suppression, the Microbubble Localisation Network (MBL-Net) for precise MB localisation, and the Microbubble Trajectory Network (MBT-Net) for MB tracking. This modular design enables the trained networks to be deployed as a plug-and-play module within conventional ULM pipelines, or combined into an end-to-end DL pipeline for fully automated ULM post-processing. Across in silico and in vivo datasets, CycleULM improves contrast, spatial resolution, localisation accuracy and super-resolution vascular reconstructions, while providing order-of-magnitude speed-ups and real-time throughput (up to 18.3 frames per second and $\sim$14.5-fold faster than conventional pipelines). Together, these advances make CycleULM a practical route toward robust, real-time and clinically translatable ULM.

\section{Results}

\begin{figure}[]
\centering
\vspace{-1cm}
\includegraphics[width=0.99\linewidth]{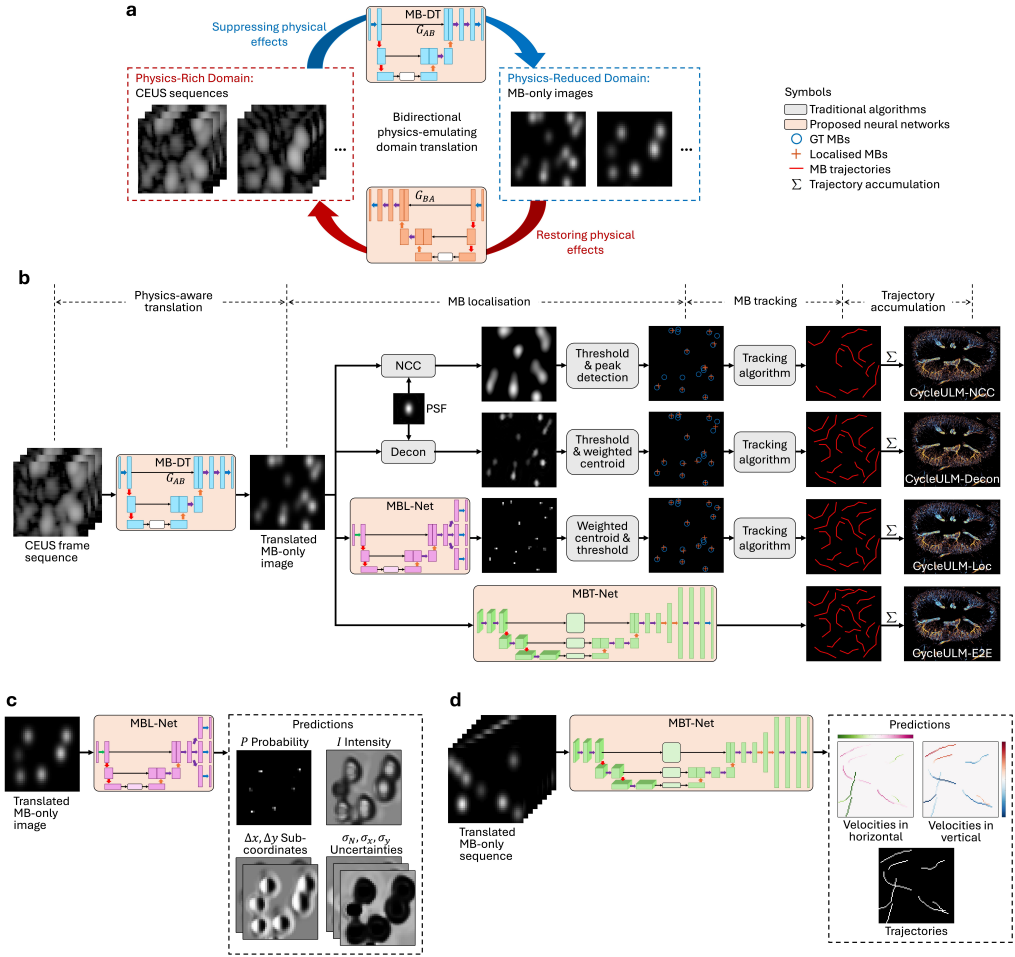}
\vspace{-0.1cm}
\caption{\textbf{Overview of the CycleULM framework.} \textbf{a}, CycleULM learns a bidirectional physics-emulating translation between the CEUS frames and a simplified simulated MB-only domain via self-supervised cycle-consistent adversarial training. The translation by MB-DT effectively bridges the domain gap between simulated data and real acquisitions, enabling the downstream processing or model training within a simplified, well-controlled domain without requiring manual annotations or complex simulators. \textbf{b}, The variant CycleULM methods. The modular design of CycleULM offers the flexibility that three networks can be integrated as plug-and-play modules within conventional ULM pipelines (CycleULM-NCC, CycleULM-Decon and CycleULM-Loc), or combined into an end-to-end DL pipeline for fully automated ULM post-processing (CycleULM-E2E). \textbf{c}, MBL-Net takes the MB-only image translated by MB-DT as input and produces four complementary outputs, including a probability map denoting the likelihood of a MB at each pixel, an intensity map estimating the bubble intensities, two sub-pixel coordinate maps that refine the accuracy of MB locations, and three uncertainty maps that quantify confidence in these predictions. \textbf{d}, MBT-Net takes a short sequence of 8 consecutive MB-only frames translated by the MB-DT as input and outputs three maps, including a trajectory-probability map indicating the likelihood of a MB path present at every pixel, a velocity map showing the horizontal velocity components, and a velocity map showing the vertical velocity components, respectively.}
\label{fig:framework}
% \vspace{-0.3cm}
\end{figure}

\subsection{CycleULM Framework}

The proposed CycleULM framework is summarised in Fig. \ref{fig:framework}. We first learn a physics-emulating bidirectional translation between the CEUS data domain and a simplified simulated MB-only domain (Fig. \ref{fig:framework}a). This translation is achieved with two neural networks, $G_{AB}$ and $G_{BA}$, trained in a self-supervised manner with unpaired data based on the CycleGAN architecture \citep{zhu2017unpaired}. This approach leverages the power of CycleGAN to learn mappings between two domains without requiring paired ground truth data. The forward generator $G_{AB}$, serving as MB-DT, translates the physics-rich domain (CEUS frames from experimental acquisition containing the imaging system response, MB nonlinear responses, heterogeneous tissue properties, noise, etc.) into the physics-reduced, simplified MB-only domain, where each frame is modelled as a linear combination of MBs with a fixed-shape PSF. In practice, MB-DT suppresses complex background noise and retains clean MB signals. The backward generator $G_{BA}$ reconstructs CEUS-like frames from the MB-only images by restoring the underlying physical effects. Together, these two networks establish a cycle-consistent, physics-emulating translation that removes reliance on high-fidelity simulators or labelled in-vivo data and aligns the training and inference domains.

To exploit the temporal dependency and increase the ability to distinguish true MB signals from clutter, MB-DT takes three consecutive CEUS frames as input. In addition, the synthetic PSF used in the MB-only domain is generated by isotropically shrinking the PSF estimated from the raw CEUS data by a factor of 0.5 in both spatial dimensions, encouraging MB-DT to produce MB-only frames with effectively smaller MBs and higher contrast and thereby facilitating downstream localisation and tracking. 

Operating in the translated MB-only domain substantially reduces the complexity of subsequent processing stages. MBL-Net (Fig.~\ref{fig:framework}c) is trained entirely on synthetic MB-only images and does not require labelled in-vivo data. Inspired by DECODE \citep{speiser2021deep}, MBL-Net takes an MB-only image as input and outputs four groups of maps: (i) a probability map denoting the likelihood of a MB at each pixel, (ii) an intensity map estimating bubble intensities, (iii) two sub-pixel coordinate offset maps that refine MB locations and (iv) three uncertainty maps that quantify confidence in these predictions. The final MB positions are decided based on these outputs. 

Similarly, MBT-Net (Fig.~\ref{fig:framework}d) is also trained entirely in the MB-only domain. It takes a short sequence of eight consecutive MB-only frames as input and outputs three maps: (i) a trajectory-probability map indicating the likelihood of MB paths at each pixel and (ii) two velocity maps representing the horizontal and vertical components of MB velocity. Combining these outputs, MBT-Net reconstructs MB trajectories and estimates local blood flow velocity and direction, enabling quantitative characterisation of microvascular dynamics.

The modular design of CycleULM offers flexible integration into existing ULM pipelines (Fig.~\ref{fig:framework}b). Combining MB-DT and MBT-Net forms the end-to-end CycleULM-E2E method, which generates SR vascular maps directly from CEUS frames. Alternatively, individual components can be used as plug-and-play modules to replace conventional algorithms. Specifically, integrating MB-DT with normalised cross-correlation (NCC) and the tracking algorithm from \cite{yan2022super} yields CycleULM-NCC; combining MB-DT with deconvolution (Decon) and the tracking algorithm produces CycleULM-Decon; and coupling MB-DT with MBL-Net and the tracking algorithm results in CycleULM-Loc. All variants are evaluated on both in silico and in vivo datasets and benchmarked against traditional ULM pipelines.

\subsection{CycleULM closes the simulation-to-reality gap}

\begin{figure}[!t]
\centering
% \vspace{-1.0cm}
\includegraphics[width=0.99\linewidth]{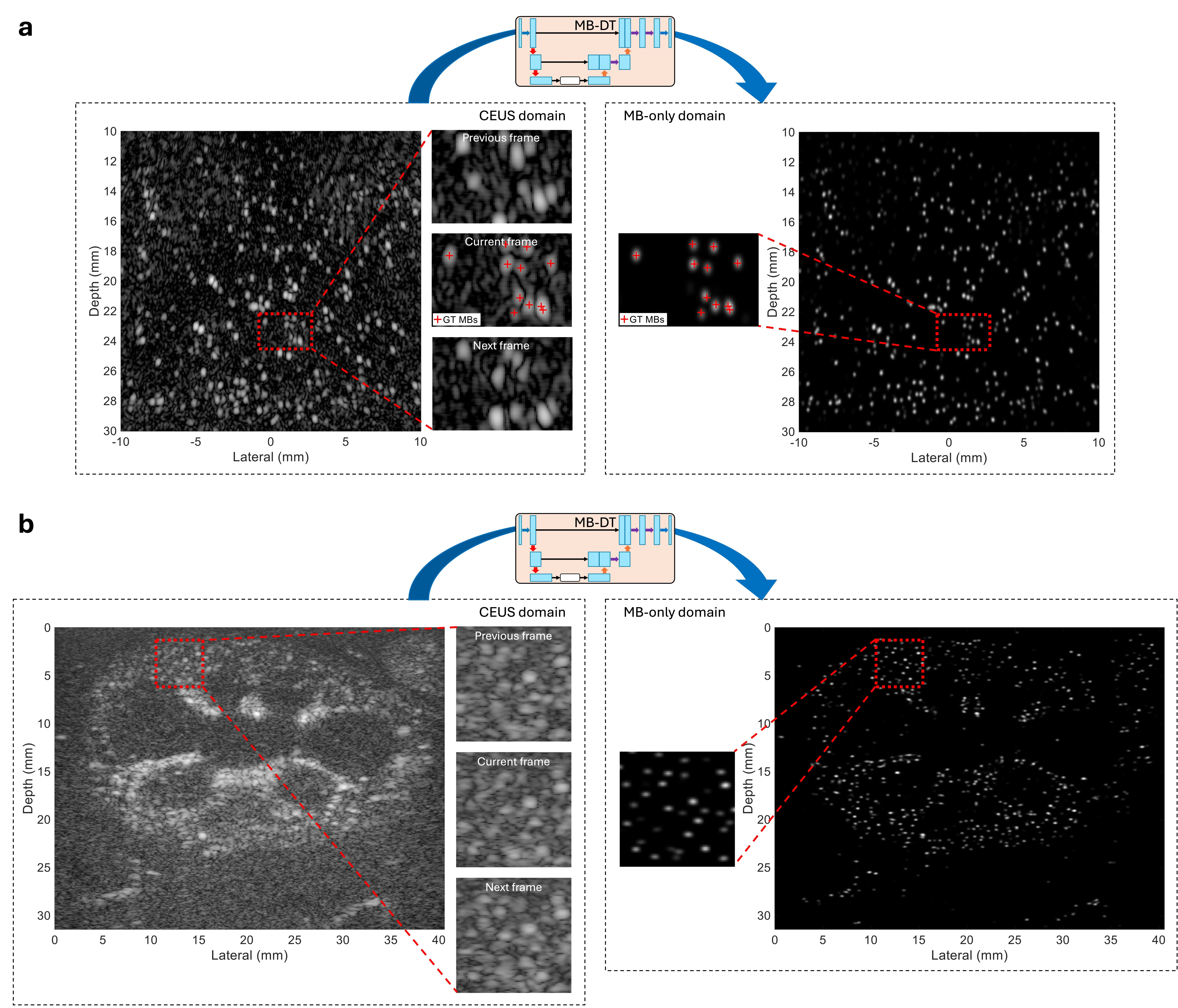}
% \vspace{-0.3cm}
\caption{\textbf{CycleULM learns a physics-emulating translation from the CEUS domain into a simplified MB-only domain.} Demonstrations on the in silico ULTRA-SR Challenge data in \textbf{a} and the in vivo rabbit kidney data in \textbf{b} show that, by leveraging the temporal consistency from three consecutive CEUS frames, the MB-DT can effectively distinguish MB signals from background clutter. This physics-emulating translation bridges the domain gap between simulated data and real acquisitions, enabling the downstream processing or model training within a simplified, well-controlled domain without requiring manual annotations or complex simulators.}
\label{fig:MB-DT}
% \vspace{-0.3cm}
\end{figure}

We first assess the ability of MB-DT to learn a physics-emulating domain translation between CEUS and the simplified MB-only domain. We use the simulated data from the Ultrasound Localisation and Tracking Algorithms for Super Resolution (ULTRA-SR) Challenge \citep{lerendegui2022bubble}, where ground truth (GT) is available for validation but not used during training. An example translation is shown in Figure~\ref{fig:MB-DT}a, with the full output sequence provided in Supplementary Video~1. MB-DT successfully converts CEUS frames into MB-only images in which background speckles are strongly suppressed and MB signals remain well preserved. Zoomed-in regions confirm that the MBs generated by MB-DT closely match the GT, demonstrating effective and robust translation.

We observe similar effects in an in vivo rabbit kidney dataset (Fig.~\ref{fig:MB-DT}b and Supplementary Video~2). The MB-only image translated by MB-DT exhibits a markedly cleaner background than the raw CEUS frame, while MB signals align closely with the original CEUS data. These results suggest that MB-DT learns to recognise MB signals from complex physical effects directly from real CEUS frames. Moreover, exploiting temporal consistency across three consecutive CEUS frames further enhances this discrimination.

By translating complex CEUS frames into a simplified, physics-reduced, MB-only domain where only MB signals with fixed-shape PSFs are present, MB-DT effectively closes the domain gap between simulated and real data. This aligns training and inference domains, eliminating the performance degradation. Downstream processing and model training are shifted from the complex CEUS domain to this controlled space, from which speckles, noise and other hard-to-model acoustic and physiological effects have been effectively removed. Instead of struggling with creating realistic CEUS simulators, we can now generate unlimited training data in the MB-only domain with known PSFs, positions and trajectories fully under control. As a result, MBL-Net and MBT-Net trained purely on the synthetic MB-only data can generalise seamlessly to real acquisitions.

\subsection{CycleULM improves image contrast and resolution}

\begin{figure}[]
\centering
\vspace{-2.6cm}
\includegraphics[width=0.9\linewidth]{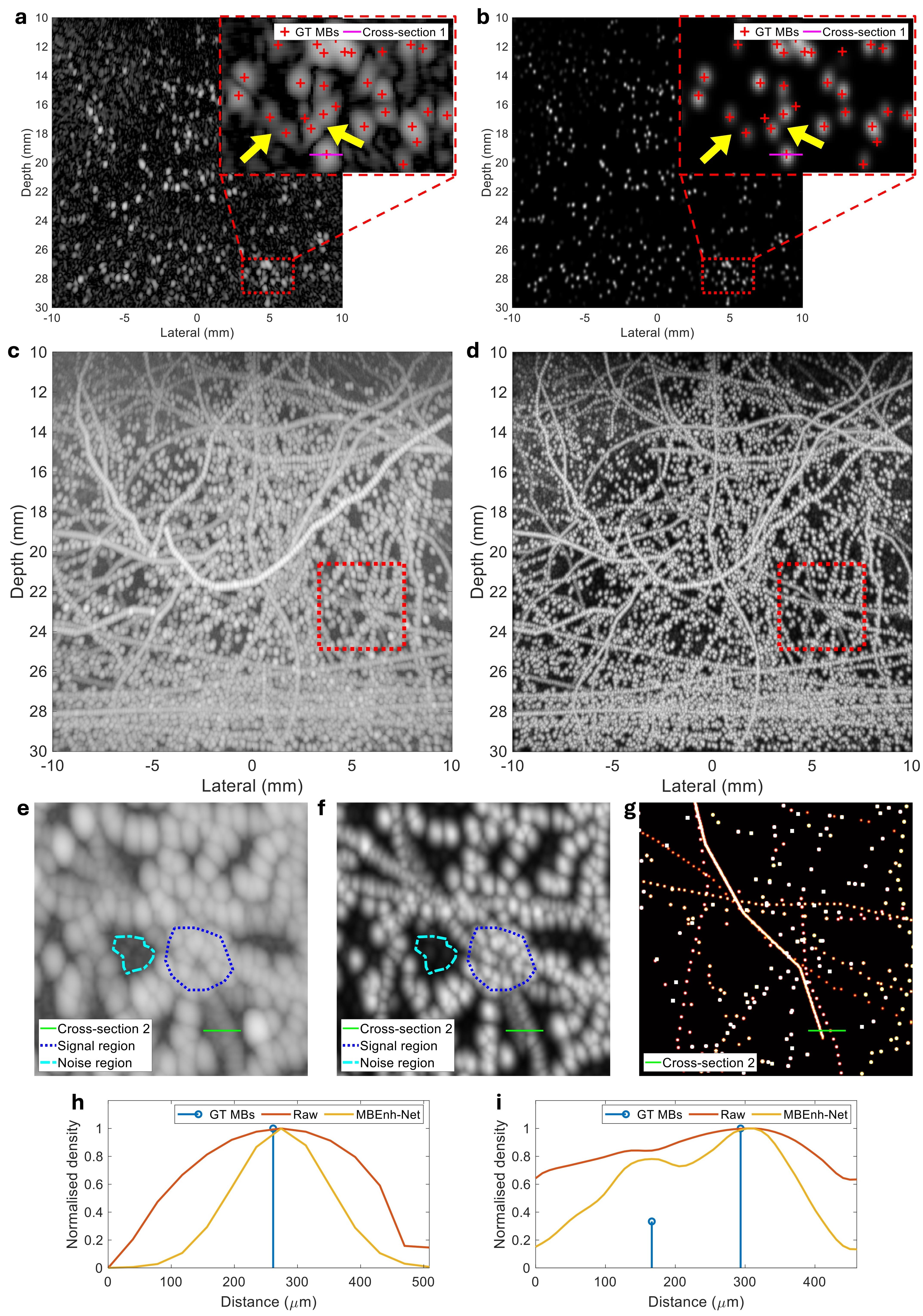}
\vspace{-0.3cm}
\caption{\textbf{CycleULM improves image contrast and resolution on in silico ULTRA-SR Challenge dataset.} \textbf{a}, An example CEUS frame and \textbf{b}, the corresponding MB-only frame translated by the MB-DT. In the MB-only frame, background clutter is significantly suppressed and the PSF size is visibly reduced, resulting in the improved separation of overlapping MBs (yellow arrows). \textbf{c}, MIPs of the raw CEUS dataset and \textbf{d}, the MB-DT output sequence. The MIP after MB-DT preserves the vascular pattern in the raw MIP but exhibits markedly lower background, higher contrast, and finer structural detail. \textbf{e,f,g}, Zoomed-in MIPs and the accumulated GT MB image show a substantial CNR improvement of 9.2 dB after MB-DT. \textbf{h}, Cross-sectional analyses across a single MB and \textbf{i}, two adjacent vessels demonstrate enhanced resolution. The measured FWHM decreased from 355 µm to 178 µm (a 2-fold resolution improvement), and overlapping MB signals became clearly separable after MB-DT.}
\label{fig:MB-DT_challenge}
% \vspace{-0.3cm}
\end{figure}

\begin{figure}[]
\centering
\vspace{-0.5cm}
\includegraphics[width=0.99\linewidth]{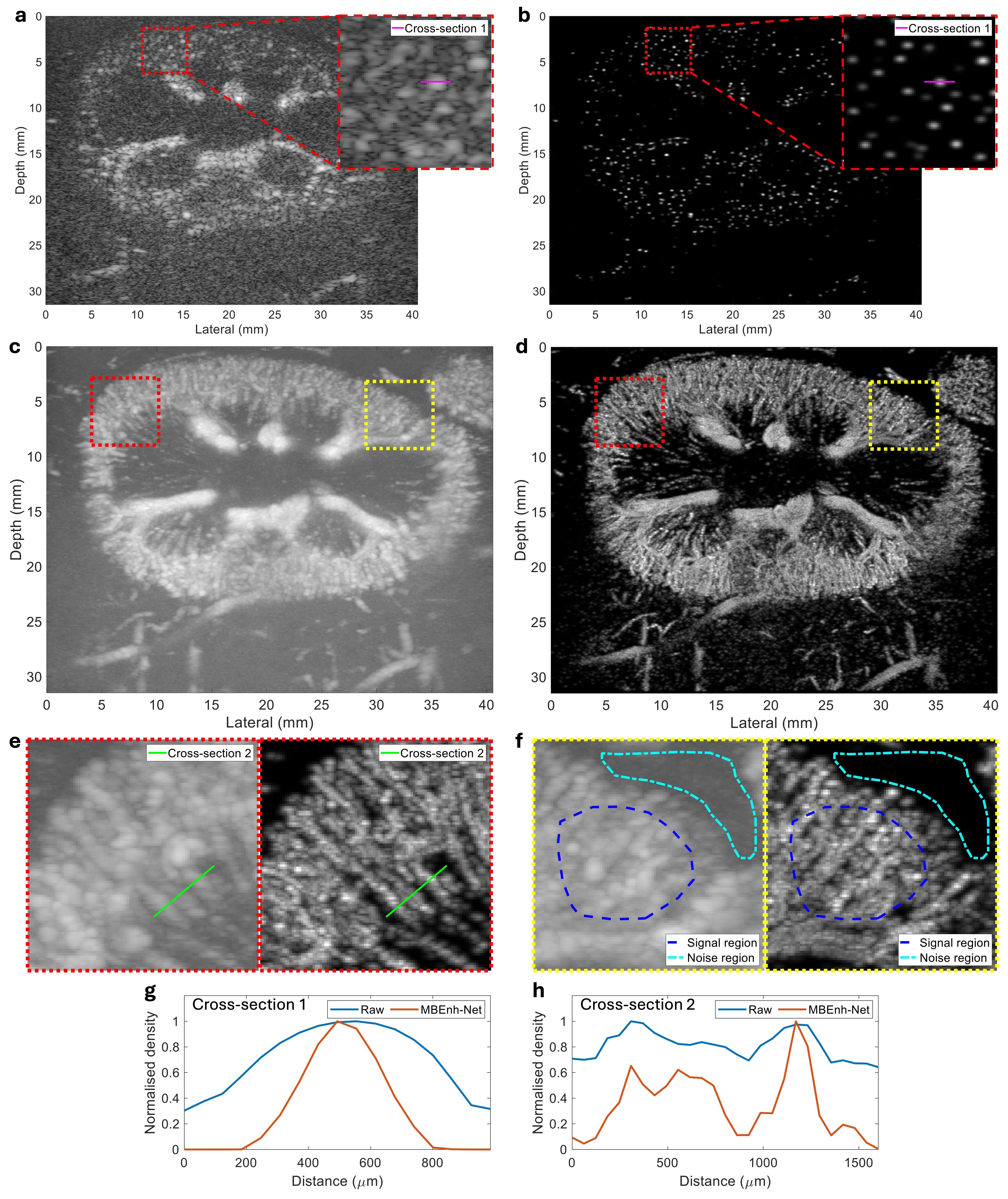}
\vspace{-0.3cm}
\caption{\textbf{CycleULM improves image contrast and resolution in an in vivo rabbit kidney dataset.} \textbf{a}, An example CEUS frame and \textbf{b}, the corresponding MB-only frame translated by the MB-DT. The MB-only shows markedly reduced background clutter. \textbf{c}, MIPs of the raw CEUS dataset and \textbf{d}, the sequence output by MB-DT. The MIP after MB-DT preserved the vascular architecture in the raw MIP while achieving substantially higher contrast and resolution. \textbf{e,f}, Zoomed-in MIPs show an extraordinary CNR improvement of 15.3 dB after MB-DT \textbf{g}, Cross-sectional analyses across a single MB and \textbf{h}, several adjacent vessels. The measured FWHM decreased from 725 µm to 294 µm after MB-DT, corresponding to a 2.5-fold resolution improvement. Adjacent vascular structures also appear more clearly separated in the MB-DT–translated MIP.}
\label{fig:MB-DT_rabbit}
% \vspace{-0.3cm}
\end{figure}

Consistent with its design, MB-DT suppresses clutter and reduces the effective PSF size, leading to improved contrast and spatial resolution in the translated MB-only images (Figs.~\ref{fig:MB-DT_challenge}, \ref{fig:MB-DT_rabbit}). On the ULTRA-SR dataset, MB-only frames produced by MB-DT exhibit strong background suppression and visibly reduced PSFs compared with the original CEUS frames, improving separation of overlapping MBs (Fig.~\ref{fig:MB-DT_challenge}a,b). The maximum intensity projection (MIP) generated after the MB-DT preserves the vascular pattern of the raw CEUS MIP while showing lower background, higher contrast and finer structural detail (Fig.~\ref{fig:MB-DT_challenge}c,d). Quantitatively, contrast enhancement was validated by the contrast-to-noise ratio (CNR) with values of 7.2 dB for the original MIP and 16.4 dB for the MIP after the MB-DT, a significant improvement of 9.2 dB (Fig. \ref{fig:MB-DT_challenge}e,f). The CNR is calculated as the ratio of the mean value of the signal region (in blue) to the mean value of the noise region (in cyan). Resolution benefits were confirmed by cross-sectional analyses along two lines: one across a single MB (magenta line) and the other across two adjacent vessels (green line). The MB width along the magenta line was significantly reduced after processing with MB-DT (\ref{fig:MB-DT_challenge}h). The measured full width at half maximum (FWHM) of the MB decreased from 355 µm in the original CEUS frame to 178 µm in the MB-DT translated MB-only image, a 2-fold improvement in resolution. In the two-MB case, the two peaks due to overlapping MB signals became clearly more separable after MB-DT (\ref{fig:MB-DT_challenge}i).

Similar effects are observed in the in vivo rabbit kidney dataset (Fig.~\ref{fig:MB-DT_rabbit}). MB-only frames produced by MB-DT show markedly reduced background speckles relative to the raw CEUS frames (Fig.~\ref{fig:MB-DT_rabbit}a,b), and the MIP after MB-DT preserves the vascular architecture while enhancing vessel delineation (Fig.~\ref{fig:MB-DT_rabbit}c,d). In the zoomed-in region, the CNR increases from 14.4~dB to 29.7~dB, showing an extraordinary 15.3 dB improvement (Fig.~\ref{fig:MB-DT_rabbit}e,f). Moreover, the FWHM of the isolated MB decreases from 725~µm to 294~µm, corresponding to a 2.5-fold resolution improvement (Fig.~\ref{fig:MB-DT_rabbit}h). Cross-sections across multiple vessels highlight improved separation of adjacent structures in the MB-DT-processed MIP (Fig.~\ref{fig:MB-DT_rabbit}i). Together, these results demonstrate that MB-DT reliably detects MB signals and substantially improves contrast and resolution in both simulated and in vivo CEUS data.

\subsection{CycleULM improves MB localisation performance}

\begin{figure}[]
\centering
\includegraphics[width=0.99\linewidth]{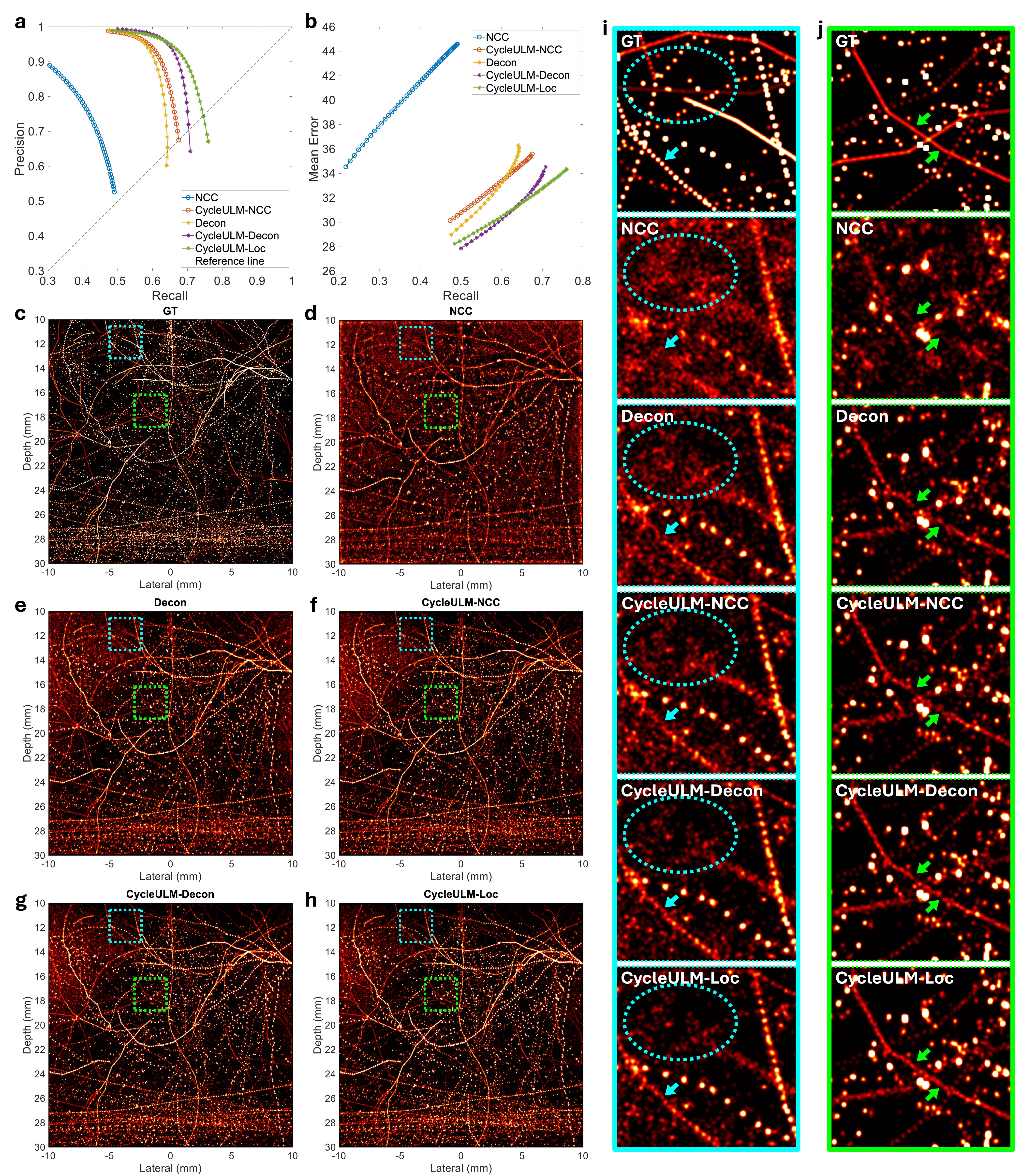}
%	\vspace{-0.3cm}
\caption{\textbf{CycleULM significantly improves MB localisation performance on in silico ULTRA-SR Challenge dataset.} \textbf{a}, Precision–recall curves. \textbf{b}, Mean localisation error versus recall curves. \textbf{c}, Accumulated GT MB image. \textbf{d–h}, Accumulated super-localised MB maps generated by different methods, each with the same number of detected MBs. \textbf{i,j}, Zoomed-in comparisons of two selected regions from the super-localised MB maps, highlighting differences in localisation accuracy and background suppression.}
\label{fig:localisation}
% \vspace{-0.3cm}
\end{figure}

Using the GT annotations provided by the ULTRA-SR Challenge dataset, we also evaluated microbubble localisation performance using the metrics described in \cite{yan2025enhancing}: 1) Recall measures the proportion of actual MBs that are correctly localised; 2) Precision quantifies the proportion of correctly localised MBs among all detected MBs; 3) $F_1$ score provides a harmonic mean of Recall and Precision, offering a balanced assessment of localisation performance. To further assess localisation accuracy, we computed the mean localisation error between the detected MB positions and the GT MB positions. By sweeping the decision thresholds across methods, we obtained precision–recall curves (Fig.~\ref{fig:localisation}a) and localisation error–recall curves (Fig.~\ref{fig:localisation}b). Notably, even when using the same conventional localisation methods (NCC and Decon), applying MB-DT–translated data led to a notable improvement in localisation performance. Specifically, compared with the Decon results on raw data, CycleULM-Decon recovered up to 7\% more true MBs at the same precision, localised MBs up to 27\% more accurately and reduced the mean localisation error by up to 4.5 µm (13\%) at the same recall. MB-DT also enhanced the performance of NCC to levels comparable to Decon on raw data. Compared with NCC, CycleULM-NCC achieved up to 32\% higher recall given the same precision level. At the same recall, CycleULM-NCC detected real MBs more accurately, with up to 46\% more precision, and reduced the mean localisation error, with up to 14.0 µm (31\%) improvement. The best MB localisation performance was obtained by the full learning-based method, CycleULM-Loc, which traced the upper envelope of the precision-recall plot while maintaining one of the lowest error-recall curves, providing the best overall trade-off between detection sensitivity and localisation accuracy. Quantitatively, CycleULM-Loc localised MBs with up to 46\% more precision, 40\% more recall and 16.1 µm (36\%) less mean localisation error compared with NCC, and up to 26\% more precision, 8\% more recall and 3.4 µm (10\%) less mean localisation error compared with Decon. 

\begin{table}[t]
\centering
\caption{\textbf{Quantitative evaluation of MB localisation performance on the in silico ULTRA-SR dataset.} The results show that our CycleULM methods with MB-DT consistently produced maps with lower MSE and higher SSIM, as well as improved recall, precision, and F1 score compared with baseline methods. Among all methods, CycleULM-Loc achieved the best overall performance (bold), followed by CycleULM-Decon (underlined), demonstrating the effectiveness of CycleULM in improving MB localisation performance.}
\includegraphics[width=1\linewidth]{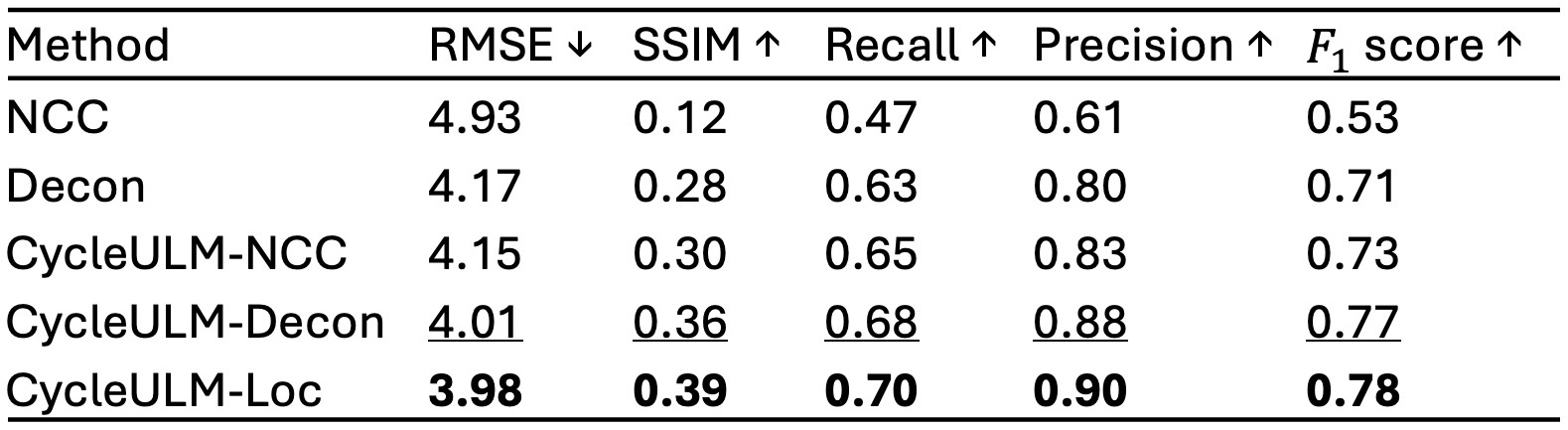}
%	\vspace{-0.3cm}
\label{tab:localisation}
\end{table}

To further assess localisation performance, we accumulate detected MBs into super-localised MB maps (Fig.~\ref{fig:localisation}d–h), ensuring that each map contains the same number of localisations for fair comparison, and benchmark them against the GT map (Fig.~\ref{fig:localisation}c). Visually, the maps generated with MB-DT exhibit better alignment with the GT, reduced background and more clearly defined structures. Two zoomed-in regions (Fig. \ref{fig:localisation}i–j) highlight these improvements: cyan circles indicate suppressed false detections, while the arrows point to additional true MBs recovered by CycleULM methods, resulting in more complete vascular features. Quantitative metrics in Table~\ref{tab:localisation} show that CycleULM methods with MB-DT consistently produced maps with lower RMSE and higher SSIM, along with improved recall, precision, and F1 score compared with baselines. Among all methods, CycleULM-Loc achieved the best overall performance, followed by CycleULM-Decon, demonstrating the effectiveness of CycleULM in enhancing MB localisation.

\subsection{CycleULM improves SR map quality}

\begin{figure}[!t]
\centering
\includegraphics[width=1\linewidth]{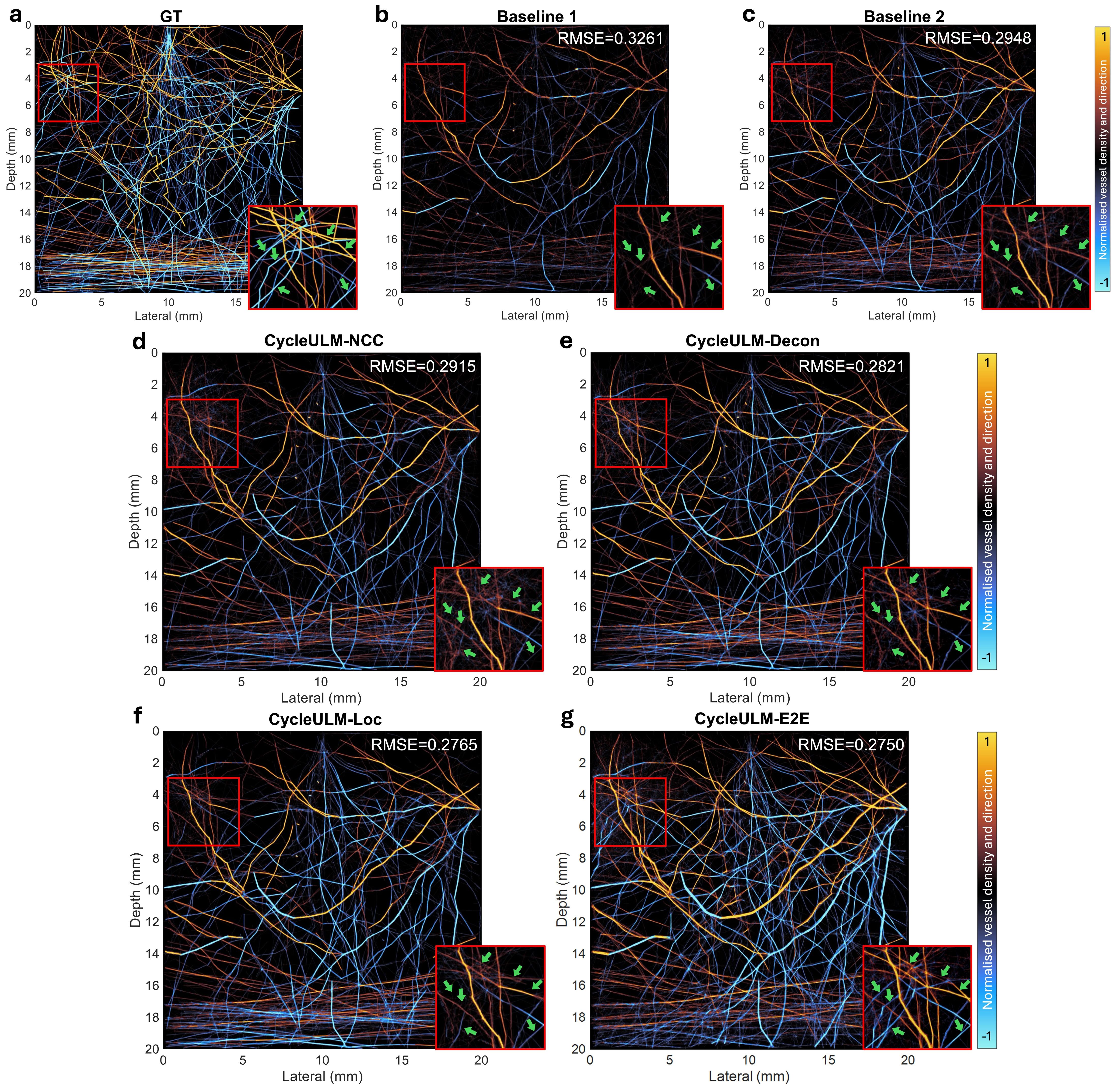}
% \vspace{-0.3cm}
\caption{\textbf{CycleULM enhances the SR ULM images on the in silico ULTRA-SR Challenge dataset.} \textbf{a}, The GT vascular map. \textbf{b,c}, The SR maps generated by the two baseline methods. \textbf{d-g}, The SR maps generated by the CycleULM methods. The CycleULM methods recovered more detailed and continuous microvascular structures compared to both baselines. In particular, CycleULM-E2E reconstructed visibly finer and more complete vessel networks (green arrows). Quantitative evaluation using RMSE supports that all CycleULM methods achieved lower RMSEs than Baseline 1 and Baseline 2, with CycleULM-E2E having the lowest RMSE of 0.2750, indicating the most accurate reconstruction of the vascular map relative to the GT.}
\label{fig:challenge_SR_maps}
% \vspace{-0.3cm}
\end{figure}

\begin{figure}[!t]
\centering
\includegraphics[width=1\linewidth]{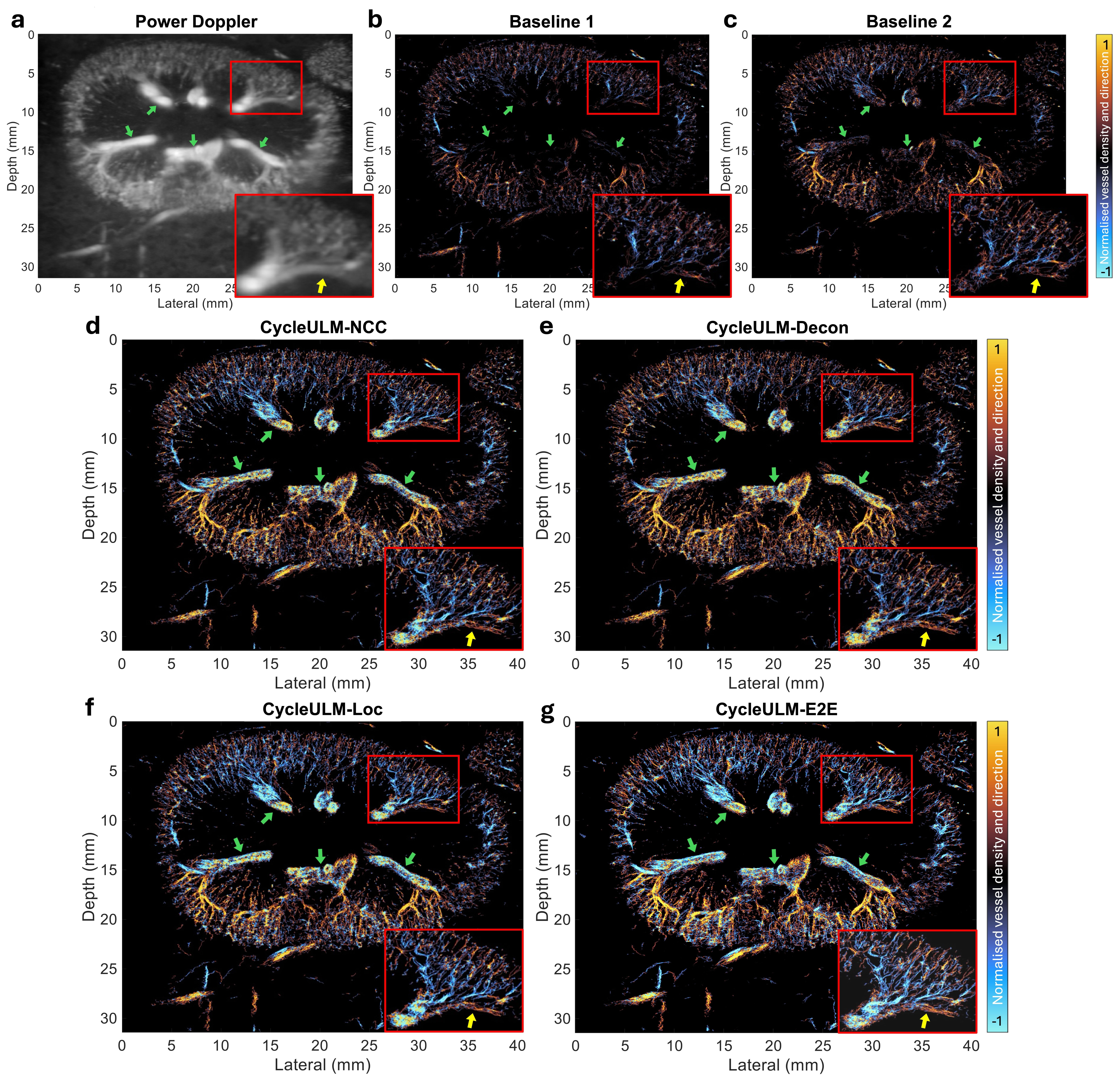}
%	\vspace{-0.3cm}
\caption{\textbf{CycleULM enhances the SR ULM images on the in vivo rabbit kidney dataset.} \textbf{a}, Power Doppler image created from 500 frames. \textbf{b,c}, The SR maps generated by the two baseline methods. \textbf{d-g}, The SR maps generated by the CycleULM methods. Compared with both baselines, the CycleULM methods recovered finer vascular branches and achieved more complete vessel reconstruction, particularly in the renal hilum areas.}
\label{fig:rabbit_SR_maps}
% \vspace{-0.3cm}
\end{figure}

We next evaluated the impact of CycleULM on the final SR vascular maps. We compared CycleULM methods against two conventional processing pipelines, Baseline~1 and Baseline~2 (Supplementary Fig.~1). Baseline~1 comprises adaptive image thresholding \citep{bradley2007adaptive} for denoising, NCC for MB localisation and the tracking algorithm in \cite{yan2022super}; Baseline~2 replaces NCC with Decon. For fair comparison, each method was constrained to produce the same total number of localised MBs, which were then linked with the same tracking algorithm, and SR images are generated by accumulating MB tracks no shorter than four frames.

On the ULTRA-SR dataset, CycleULM methods clearly recover more detailed and continuous microvascular structures than both baselines (Fig.~\ref{fig:challenge_SR_maps}). In particular, CycleULM-E2E reconstructs the finest and most complete vessel networks (green arrows), aligning the best with the GT vascular map (Fig.~\ref{fig:challenge_SR_maps}a). RMSE values computed between reconstructed SR images and the GT confirm these observations: all CycleULM methods achieve lower RMSE than baselines, with CycleULM-E2E attaining the lowest RMSE of 0.2750.

\begin{figure}[!t]
\centering
\includegraphics[width=1\linewidth]{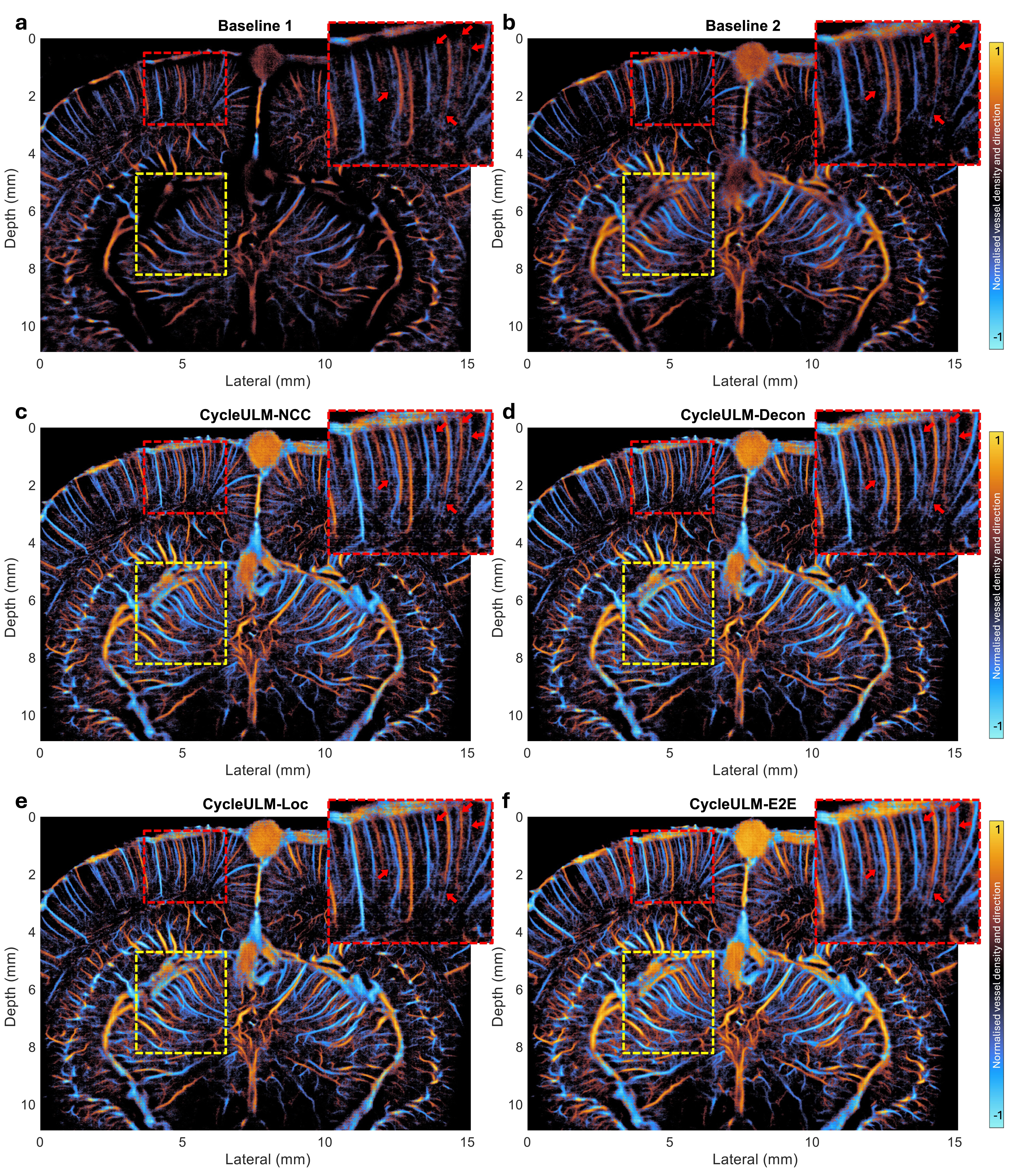}
%	\vspace{-0.3cm}
\caption{\textbf{CycleULM enhances the SR ULM images in an in vivo rat brain dataset.} \textbf{a,b}, The SR maps generated by the two baseline methods. \textbf{c-f}, The SR maps generated by the CycleULM methods. Compared with both baselines, the CycleULM methods recovered finer and more complete cerebral vasculature reconstruction, highlighted by coloured boxes and arrows.}
\label{fig:new_brain}
% \vspace{-0.3cm}
\end{figure}

Figure~\ref{fig:rabbit_SR_maps} compares the power Doppler image with SR vascular maps generated using different methods on the in vivo rabbit kidney dataset, comprising 500 frames equivalent to a 5-second acquisition. Baseline~1 generates limited vessel reconstruction in the renal hilum region (green arrows), despite strong MB presence in the power Doppler image, reflecting the limited ability of NCC to localise and separate overlapping MBs. Baseline 2 with Decon method is much better at separating MBs, recovering more detailed vessels in this area. More importantly, when both conventional localisation methods are applied to MB-DT translated frames (CycleULM-NCC and CycleULM-Decon), the resulting SR maps show substantial improvements, revealing finer details in the renal hilum and more clearly defined vascular tree structures. The zoomed-in region highlights that CycleULM methods reconstruct finer vessel structures and more complete vessel trees, especially for the end-to-end deep-learning approach, CycleULM-E2E. 

Similar improvements are observed in Figure~\ref{fig:new_brain}, which shows SR ULM images generated on a publicly available rat brain dataset from~\cite{shin2024context}, comprising 25000 frames acquired in 100 seconds. Compared with the two baselines, the CycleULM methods recover a finer and more complete depiction of the cerebral microvasculature, again with the highest image quality achieved by CycleULM-E2E. Together, these results demonstrate the effectiveness and robustness of the proposed CycleULM framework across different organs and acquisition conditions.

\subsection{CycleULM accelerates ULM processing to real-time}

\begin{table}[t]
\centering
\caption{\textbf{CycleULM speeds up the ULM processing to real-time.} Processing times by different methods in seconds and their throughputs in frames per second. CycleULM methods are all faster than the two baseline methods. Notably, the end-to-end deep learning model, CycleULM-E2E (underlined), achieved the fastest SR image generation, requiring only 27.2 seconds for 500 frames, reaching a processing throughput of 18.4 frames per second. Each frame had a spatial size of 511 × 659 pixels and was processed with an upsampling factor of 4. Even without speed-specific optimisations, these results demonstrate the strong potential of CycleULM for real-time ULM imaging.}
\includegraphics[width=1\linewidth]{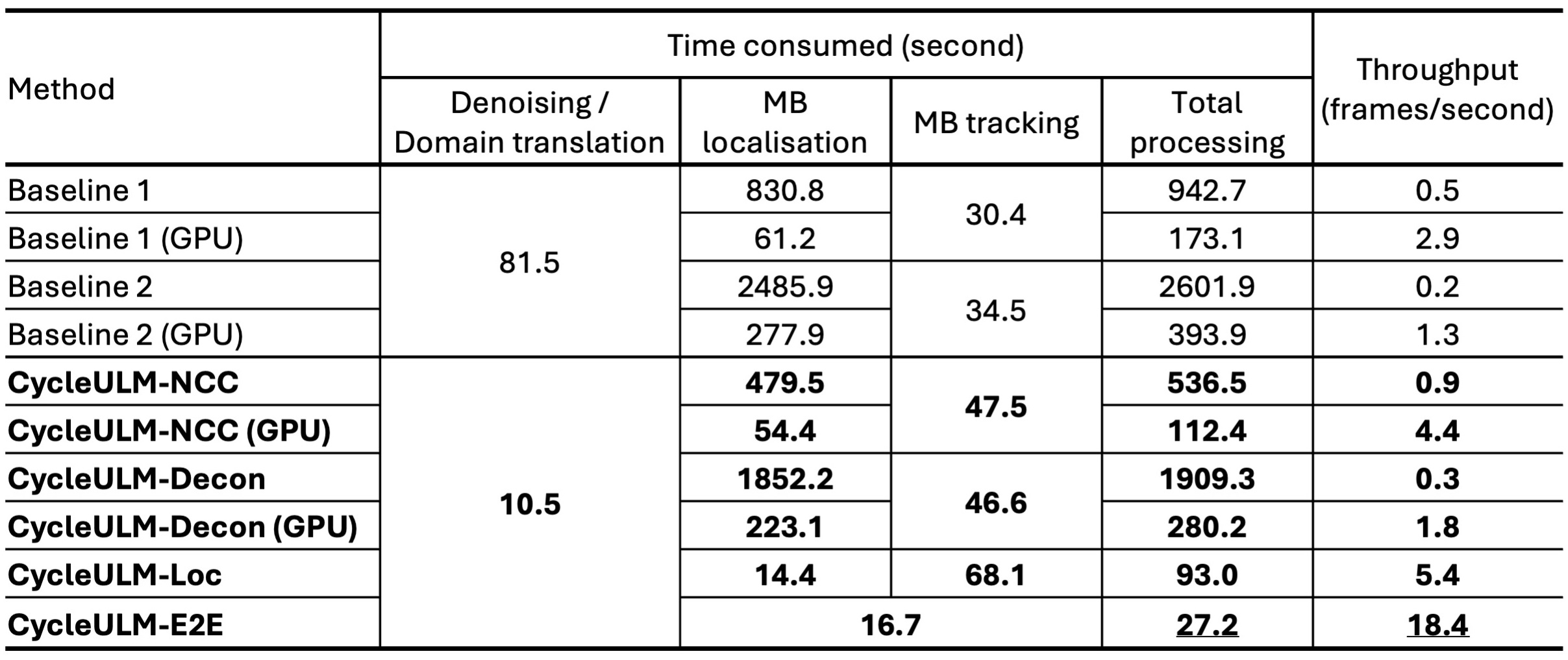}
%	\vspace{-0.3cm}
\label{tab:rabbit_SR_time}
\end{table}

Speed is another key advantage provided by CycleULM. We evaluated the computational efficiency of CycleULM on the in vivo rabbit kidney dataset, having 500 frames, each with a spatial resolution of 511$\times$659 pixels, and the final SR images were upsampled by a factor of 4 in both spatial dimensions during processing. All methods were implemented and run on the same GPU to ensure a fair comparison. Table~\ref{tab:rabbit_SR_time} summarises the processing times and throughputs.

MB-DT substantially reduces denoising time from 81.5~s for adaptive thresholding to 10.5~s, a 7.8-fold speed-up. Localisation is also accelerated: because MB-DT shrinks the effective PSF by a factor of 0.5, the computational burden of NCC and Decon is reduced, making CycleULM-NCC and CycleULM-Decon up to approximately 1.5–1.8 times faster than their raw-CEUS counterparts on both CPU and GPU. Tracking time increases slightly for CycleULM variants, likely because they localise more true MBs, leading to more trajectories to link.

Among all methods, the end-to-end deep-learning method, CycleULM-E2E, is the fastest, requiring only 27.2 seconds to generate an SR image from 500 frames, corresponding to a processing throughput of 18.4 frames per second. This represents speed-ups of 6.4-fold over Baseline 1 and 14.5-fold over Baseline 2, even though all methods are GPU-accelerated. These results demonstrate that CycleULM not only improves image quality and localisation performance, but also accelerates ULM processing to a level that potentially allows implementation.

\subsection{CycleULM generalises to independent rabbit kidney acquisitions}

To assess whether CycleULM generalises beyond the specific acquisition used for training, we evaluated the framework on an independent in vivo rabbit kidney dataset acquired using a different imaging view. In this experiment, MB-DT was trained exclusively on the first rabbit kidney acquisition (in Fig. \ref{fig:MB-DT_rabbit}) and then applied directly to the new acquisition without any fine-tuning. The localisation and tracking networks (MBL-Net and MBT-Net) were also identical to those used in the previous rabbit kidney experiment.

\begin{figure}[!t]
\centering
\includegraphics[width=1\linewidth]{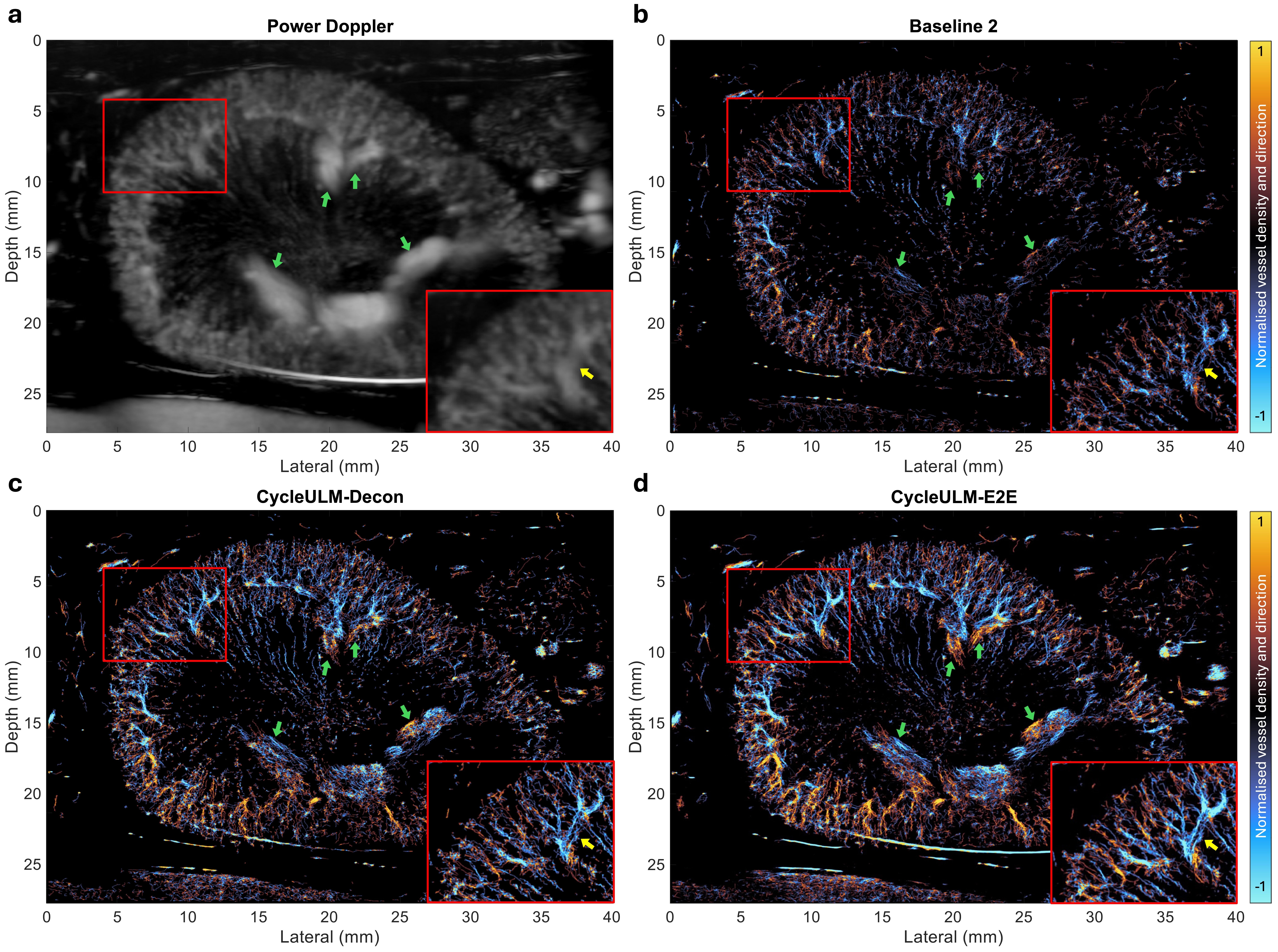}
%	\vspace{-0.3cm}
\caption{\textbf{CycleULM enhances the SR ULM images on the in vivo rabbit kidney dataset.} \textbf{a}, Power Doppler image created from 500 frames. \textbf{b}, The SR map generated by Baseline 2 method. \textbf{c,d}, The SR maps generated by CycleULM-Decon and CycleULM-E2E methods. Compared with both baselines, the CycleULM methods recovered finer vascular branches and achieved more complete vessel reconstruction, particularly in the renal hilum areas.}
\label{fig:new_rabbit_SR_maps}
% \vspace{-0.3cm}
\end{figure}

Fig.~\ref{fig:new_rabbit_SR_maps}a,b show the power Doppler image and the corresponding SR vascular map obtained with Baseline~2 on the new dataset. As in the first acquisition, Baseline~2 recovers only limited vessel structures, with particularly sparse reconstruction in the renal hilum region where MBs are present at a high concentration. In contrast, CycleULM-Decon (Fig.~\ref{fig:new_rabbit_SR_maps}c) and CycleULM-E2E (Fig.~\ref{fig:new_rabbit_SR_maps}d) still reveal finer details in the hilum and more clearly defined vascular trees, with CycleULM-E2E providing the sharpest and most complete reconstruction, as highlighted in the zoomed-in region.

These results demonstrate that the networks in CycleULM trained on a single rabbit kidney acquisition can be applied directly to an independent acquisition with a different imaging view, without retraining or parameter adjustment, while preserving their enhancements in the resulting SR microvascular maps. This cross-acquisition experiment supports the generalisability and robustness of CycleULM.

\section{Discussion}

In this paper, we present CycleULM, a unified deep-learning framework for ULM that achieves high-quality SR vascular imaging without requiring labelled in vivo data or high-fidelity simulators. CycleULM comprises three networks: the Microbubble Domain Translator (MB-DT), the Microbubble Localisation Network (MBL-Net) and the Microbubble Trajectory Network (MBT-Net). At its core, MB-DT enables self-supervised training of a bidirectional, physics-emulating translation between acquired CEUS frames and simulated MB-only images without the need of paired data ground truth. This closes the simulation-to-reality domain gap that limits supervised DL methods and obviates the need for complex physics simulators \citep{rauby2024deep}. Within this simplified domain, MBL-Net performs uncertainty-aware, sub-pixel MB localisation using multi-head predictions of presence, intensity, coordinates and uncertainties, trained with a probabilistic loss that encourages accurate and calibrated detections from purely synthetic MB-only data. Moreover, MBT-Net exploits spatiotemporal encoding to link MBs across frames and directly estimate trajectory probabilities and velocity fields.

The effectiveness of CycleULM was validated on both in silico and in vivo datasets. On the ULTRA-SR data, MB-DT reliably suppresses background and shrinks the effective PSF, leading to significant enhancements in contrast and spatial resolution and enabling conventional localisation methods to recover more true MBs with higher accuracy and lower localisation error. The CycleULM-Loc with the MBL-Net further improves localisation performance, achieving state-of-the-art precision and recall. Together with MBT-Net, MB-DT forms an end-to-end deep learning solution, CycleULM-E2E, which generates SR vascular maps directly from CEUS frames. From both in silico and rabbit kidney data, CycleULM reconstructs finer and more continuous vessel networks than conventional pipelines and significantly improves speed, with the end-to-end CycleULM-E2E reaching real-time throughput.

Several key factors contribute to the excellent performance of our CycleULM framework. First, CycleULM exploits temporal context at multiple stages. MB-DT processes three consecutive CEUS frames, improving its ability to discriminate true MB signals from complex background, while MBT-Net integrates ConvLSTM blocks to capture and propagate temporal features during trajectory estimation. Second, the network architectures are designed to be shallow and task-specific. MBL-Net extends a U-Net backbone with residual bottlenecks and multi-head outputs for probabilities, intensities, coordinates and uncertainties, and MBT-Net inserts ConvLSTM modules in the bottleneck and skip connections to leverage temporal dependencies in multi-scale features. Third, the loss functions are designed to guide learning toward physically meaningful solutions. A similarity loss regularises MB-DT, encouraging MB-only images to remain consistent with the original CEUS frames, preventing spurious MB detections at implausible locations and reducing the risk of missing true MBs. For localisation and tracking, probability-weighted losses focus networks on regions where true MBs are likely present, improving robustness in sparse and noisy cases. Finally, all networks are trained on small patches rather than full frames. This patch-based training increases effective data diversity, promotes generalisation and enables targeted use of regions of interest, while fully convolutional architectures ensure that learned representations transfer directly to full-image inference.

Another distinctive feature of CycleULM is its modular design, which mirrors classical ULM post-processing pipelines (Supplementary Fig. 1). The overall problem is decomposed into three well-defined subtasks, image denoising, localisation and tracking, each addressed by a dedicated network. This decomposition allows each network to remain compact and shallow, thereby reducing training data requirements, accelerating convergence, avoiding overfitting and ensuring the computational efficiency of CycleULM. It also provides flexibility in both training and inference. For a new dataset, the networks can be trained in parallel, and in inference, MB-DT and MBL-Net can be used as plug-and-play replacements for the conventional algorithms in the ULM post-processing pipeline, substantially improving localisation performance and the quality of SR microvascular maps. Alternatively, combining MB-DT and MBT-Net forms an end-to-end deep-learning solution, CycleULM-E2E, which produces SR vascular images directly from CEUS frames and already shows strong potential for real-time clinical implementation even without any speed-specific optimisations.

More broadly, CycleULM addresses a central obstacle for deep learning in ULM: the domain gap between simulated training data and real acquisitions, which remains a major barrier to reliable in vivo deployment \citep{rauby2024deep}. Existing strategies to reduce this gap typically involve (i) increasing simulator complexity, (ii) labelling real data or (iii) training on pseudo-labels from conventional algorithms. More complex simulators are difficult to design and computationally expensive; manual labelling is labour-intensive and impractical at scale; and pseudo-labels inevitably inherit the errors and biases of the conventional algorithms, ultimately limiting model performance. In contrast, CycleULM highlights a new strategy that closes the simulation-to-reality domain gap. The underlying idea could, in principle, be extended to volumetric (3D) ULM and other localisation-based microscopies such as single-molecule localisation microscopy (SMLM), providing a general route to deploy supervised models without exhaustive real-data labelling.

One limitation of CycleULM is that the end-to-end CycleULM-E2E produces image-based velocity and trajectory representations rather than explicit MB tracks, as returned by conventional tracking algorithms. This limits direct access to point-wise trajectories and can complicate downstream analyses that require track objects, such as track-level filtering, lifetime statistics, or vessel-wise flow quantification. In future work, this could be addressed by adding a lightweight post-hoc track-decoding step that fuses CycleULM-Loc detections with CycleULM-E2E motion priors to reconstruct explicit MB trajectories. 

\begin{figure}[]
\centering
\includegraphics[width=0.8\linewidth]{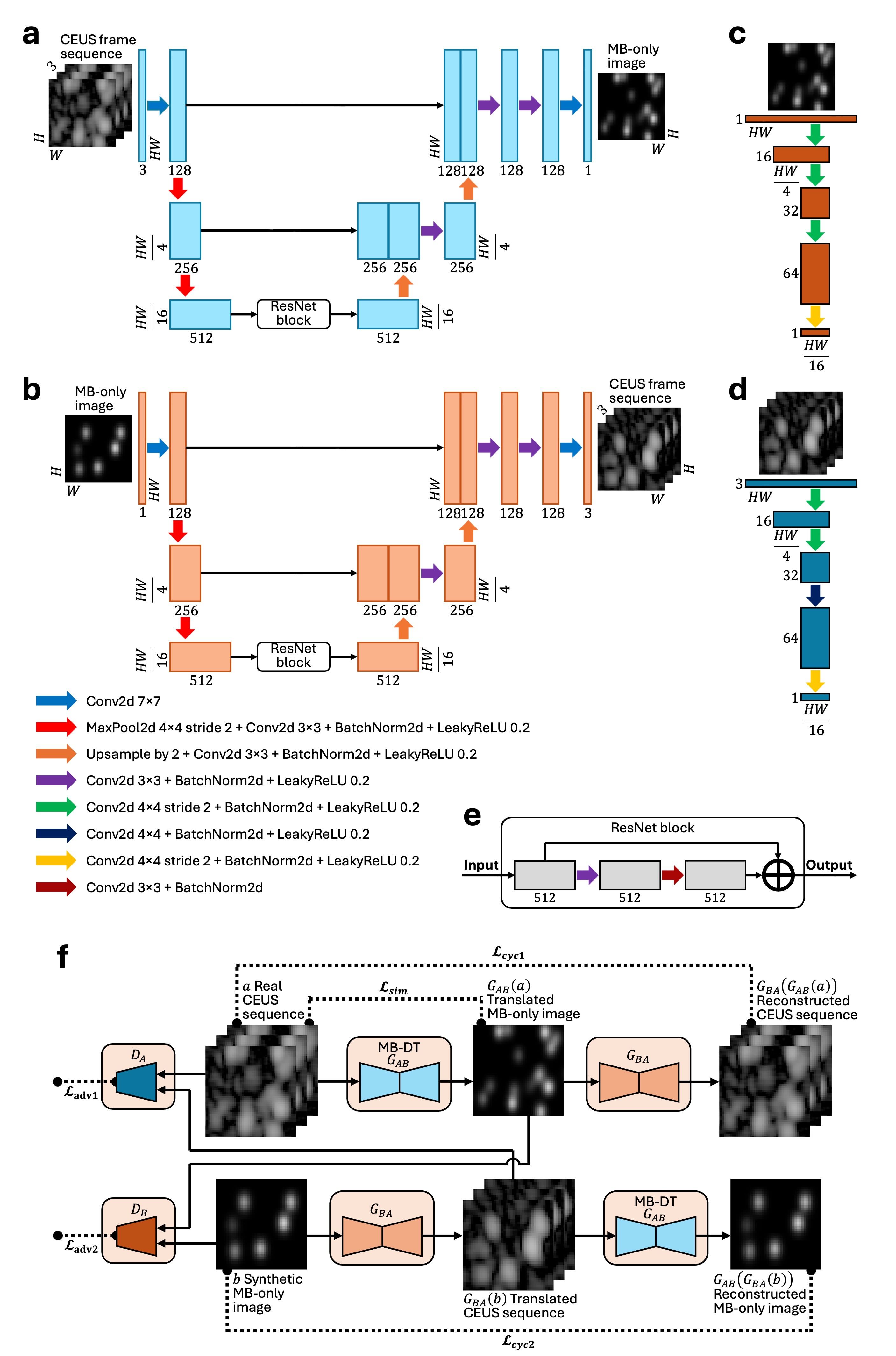}
%	\vspace{-0.3cm}
\caption{\textbf{Architectures of the Cycle-GAN and training losses for the MB-DT.} \textbf{a,b}, Network architectures of generators $G_{AB}$ (MB-DT) and $G_{BA}$, respectively. \textbf{c,d}, Network architectures of discriminators $D_A$ and $D_B$, respectively. \textbf{e}, Structure of residual network (ResNet) block used within the generators to enhance feature representation in the deeper layers. \textbf{f}, Overall training strategy and associated loss functions for learning the physics-emulating domain translation.}
\label{fig:MBE_Net}
% \vspace{-0.3cm}
\end{figure}

\section{Methods} 
\subsection{Network architectures and loss functions}
\textbf{MB-DT details.} Our MB-DT is designed based on an unsupervised generative model, Cycle-GAN \citep{zhu2017unpaired}, which consists of four networks, including two generators ($G_{AB}$ and $G_{BA}$) and two discriminators ($D_A$ and $D_B$). Both generators share a nearly identical architecture based on U-Net \citep{ronneberger2015u}, with the only distinction being their respective inputs and outputs. The detailed architectures of the generators and discriminators are provided in Fig. \ref{fig:MBE_Net}. The forward generator $G_{AB}$, which serves as MB-DT, is designed to translate multiple consecutive CEUS frames (three frames in this paper) into the simplified MB-only domain. In contrast, the backward generator $G_{BA}$ maps a MB-only image back to a sequence of three consecutive CEUS frames. The use of multiple consecutive CEUS frames enables the network to exploit temporal information across frames, enhancing MB signal extraction and clutter suppression. Following the U-Net architecture, the generators have two downsampling layers in the encoder and two upsampling layers in the decoder. Downsampling is performed using 4×4 max pooling with a stride of 2, while upsampling is implemented by bilinear interpolation followed by a 2D convolutional layer with a 3×3 kernel. To enhance feature representation in the deeper layers, we replace the original U-Net bottleneck with a residual network (ResNet) block \citep{he2016deep} consisting of two convolutional layers. This modification promotes more effective feature learning and helps prevent the loss of MB signals during network processing. The two discriminators, having the same architecture shown in Fig. \ref{fig:MBE_Net}, are designed based on the PatchGAN discriminator proposed by \cite{isola2017image}. Unlike a standard GAN discriminator, which evaluates each image by a single real or fake score, the PatchGAN discriminator classifies each image with a grid of real or fake scores. This enables it to capture high-frequency fine details and provides more efficient and stable training.

Training the Cycle-GAN involves joint optimisation of the two generators and two discriminators. The overall objective aims to minimise a combination of two cycle consistency losses, two adversarial losses, and one additional similarity loss. For more realistic outputs and more stable convergence, we implement the adversarial losses with the least-squares GAN (LSGAN) formulation \citep{mao2017least}. Accordingly, the loss used to train the generators $G_{AB}$ and $G_{BA}$ is summarised as:
\begin{align}
    \mathcal{L}_{G} = \mathcal{L}_{G_{AB}}+\mathcal{L}_{G_{BA}}+\lambda_1\mathcal{L}_{\text{cyc1}}+\lambda_2\mathcal{L}_{\text{cyc2}}+\lambda_3\mathcal{L}_{\text{sim}},
\end{align}
where $\mathcal{L}_{G_{AB}}$ and $\mathcal{L}_{G_{BA}}$ are two generator losses formulated as:
\begin{align}
    \mathcal{L}_{G_{AB}} &= \frac{1}{2} \mathbb{E}_{a \sim p_{A}(a)}\left[(D_B(G_{AB}(a)) - 1)^2\right],\\
    \mathcal{L}_{G_{BA}} &= \frac{1}{2} \mathbb{E}_{b \sim p_{B}(b)}\left[(D_A(G_{BA}(b)) - 1)^2\right],
\end{align}
$\mathcal{L}_{\text{cyc1}}$ and $\mathcal{L}_{\text{cyc2}}$ are two consistency losses expressed as:
\begin{align}
    \mathcal{L}_{\text{cyc1}} &= \mathbb{E}_{a \sim p_{A}(a)}\left[\|G_{BA}(G_{AB}(a)) - a\|_1\right], \\
    \mathcal{L}_{\text{cyc2}} &= \mathbb{E}_{b \sim p_{B}(b)}\left[\|G_{AB}(G_{BA}(b)) - b\|_1\right],
\end{align}
$\mathcal{L}_{\text{sim}}$ is the similarity loss. This additional self-supervision loss encourages the translated MB-only images to remain closely aligned with the original CEUS images, helping to preserve MB signals and reducing the risk of missing MBs during domain translation. It is defined as:
\begin{align}
    \mathcal{L}_{\text{sim}} = \mathbb{E}_{a \sim p_{A}(a)}\left[\|G_{AB}(a) - a\|_1\right],
\end{align}
and $\lambda_1$, $\lambda_2$ and $\lambda_3$ are weights that control the importance of the corresponding loss terms. They are set to be $\lambda_1=\lambda_2=5$ and $\lambda_3=1$ in this paper. The term $a \sim p_{A}(a)$ denotes three-frame image sequences sampled from the original CEUS image domain (domain A), while $b \sim p_{B}(b)$ denotes single-channel images sampled from the clutter-free MB image domain (domain B). In this paper, we set $\lambda_1=\lambda_2=5$ and $\lambda_3=1$. Then, the loss used to train the discriminators $D_A$ and $D_B$ is summarised as:
\begin{align}
    \mathcal{L}_{D} = \mathcal{L}_{D_A} + \mathcal{L}_{D_B},
\end{align}
where $\mathcal{L}_{D_A}$ and $\mathcal{L}_{D_B}$ are two discriminator losses formulated as:
\begin{align}
    \mathcal{L}_{D_A} &= \frac{1}{2} \mathbb{E}_{a \sim p_{A}(a)}\left[(D_A(a) - 1)^2\right] + \frac{1}{2} \mathbb{E}_{b \sim p_{B}(b)}\left[(D_A(G_{BA}(b)))^2\right], \\
    \mathcal{L}_{D_B} &= \frac{1}{2} \mathbb{E}_{b \sim p_{B}(b)}\left[(D_B(b) - 1)^2\right] + \frac{1}{2} \mathbb{E}_{a \sim p_{A}(a)}\left[(D_B(G_{AB}(a)))^2\right].
\end{align}

The generators and discriminators are trained on small image patches but applied to full-frame images during inference. The CEUS patches are randomly cropped from real CEUS datasets, whereas the synthetic MB-only images are created by convolving the PSF with a random number of simulated MBs placed at random locations and assigned random intensities drawn from uniform distributions. Since the MB-DT translates CEUS frames into the simplified MB-only domain, both the MBL-Net and MBT-Net are trained exclusively on synthetic MB-only patches. Once trained, these networks can be directly applied to real CEUS data following domain translation by the MB-DT, ensuring seamless generalisation from synthetic to real data without requiring labelled CEUS frames.

\begin{figure}[]
\centering
\includegraphics[width=1\linewidth]{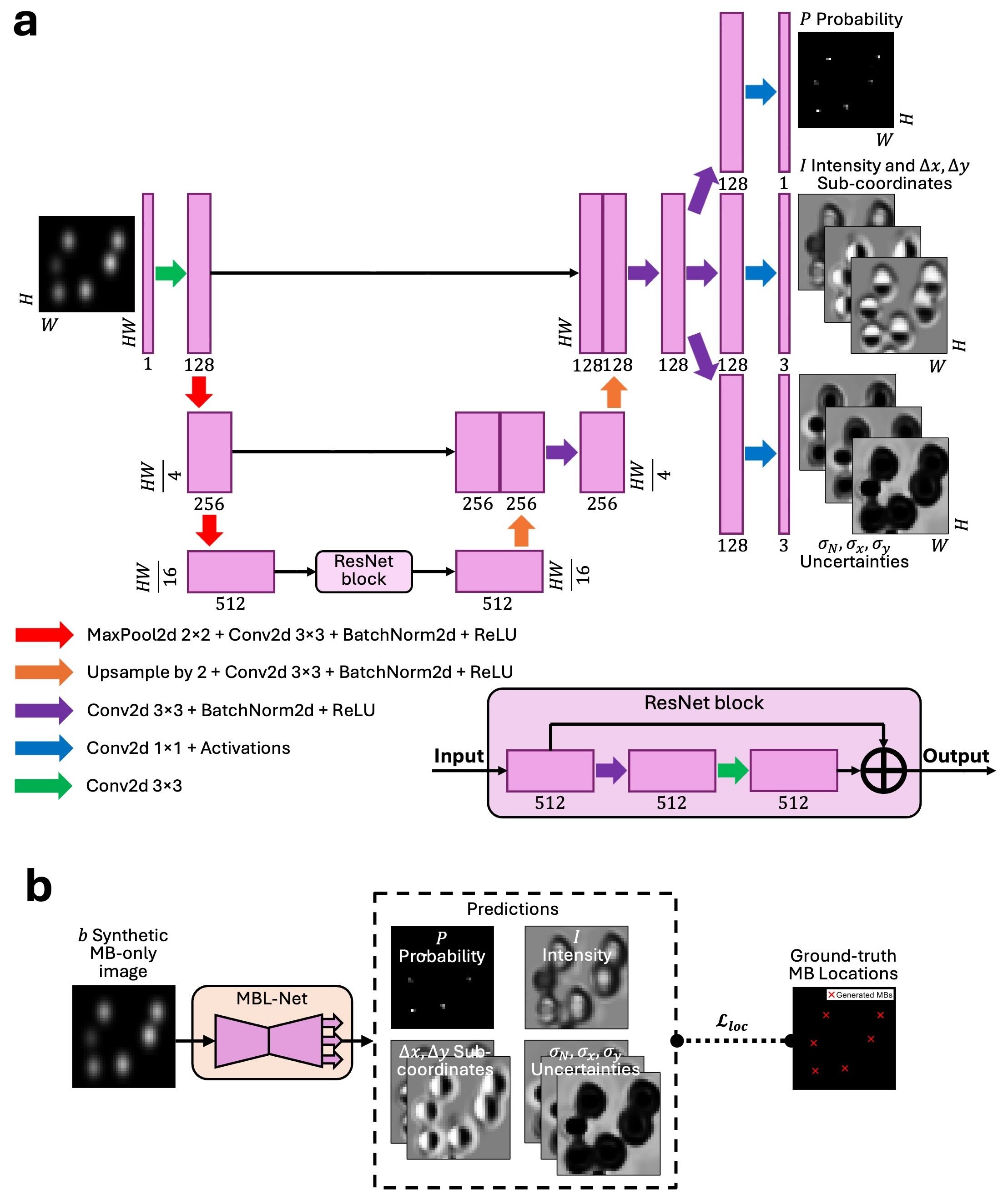}
%	\vspace{-0.3cm}
\caption{\textbf{Architecture and training losses of the MBL-Net.} \textbf{a}, Network architecture of the MBL-Net, including the structure of the residual (ResNet) block used within the network. \textbf{b}, Overall training strategy and corresponding loss functions employed for learning MB localisation in the MB-only domain.}
\label{fig:MBL_Net}
% \vspace{-0.3cm}
\end{figure}

\textbf{MBL-Net details.} Our MBL-Net is also designed based on the U-Net architecture, as shown in Fig. \ref{fig:MBL_Net}. Its main architecture is similar to the MB-DT, having two downsampling layers in the encoder and two upsampling layers in the decoder. However, there are two key differences. First, MBL-Net outputs seven prediction maps through three multi-task heads appended to the end of the U-Net, enabling the use of a powerful localisation loss inspired by the DECODE network proposed in \cite{speiser2021deep}. The seven prediction maps include: 1) a probability map $p$, indicating the likelihood of a MB being detected at each pixel; 2) an intensity map $I$, representing the MB intensity at a MB is detected at a certain pixel; 3) two sub-pixel coordinate offset maps, $\Delta x$ and $\Delta y$, which refine the MB location within the pixel grid; and 4) three corresponding uncertainty maps, $\sigma_I$, $\sigma_x$, and $\sigma_y$, quantifying the confidence of predictions. Second, MBL-Net employs smaller convolutional kernel sizes and a more refined downsampling strategy: 2×2 max pooling with a stride of 2. This configuration is chosen to improve the spatial precision and enhance the localisation accuracy of MBs.

To train the MBL-Net, we propose to minimise an overall loss that consists of two terms: a presence loss $\mathcal{L}_{\text{p}}$ and a localisation loss $\mathcal{L}_{\text{loc}}$:
\begin{align}
    \mathcal{L}_{\text{MBL}} &= \mathcal{L}_{\text{p}} + \mathcal{L}_{\text{loc}}.
\end{align}
The presence loss $\mathcal{L}_{\text{p}}$ is designed as a Binary‑Cross‑Entropy (BCE) loss between the predicted probability map $p$ and the GT binary MB presence map $p^{\text{GT}}$, expressed as:
\begin{align}
    \mathcal{L}_{\text{p}}=
-\left[p^{\text{GT}}\log p + (1-p^{\text{GT}})\log(1-p)\right].
\end{align}
This loss is log-likelihood–based and gradient-friendly, encouraging the network to produce well-calibrated probabilistic outputs and effectively estimate the likelihood of a MB being present at each pixel. The localisation loss used in this paper is a two-dimensional adaptation of the loss proposed in \cite{speiser2021deep}. Instead of directly predicting the intensity and coordinates of the $m$-th MB, represented by $\textbf{u}_m=[I_m,x_m,y_m]$, the network predicts a three-dimensional Gaussian distribution at each pixel $k$, given by:
\begin{align}
    \mathcal{P}\left(\textbf{u}_m | \bm{\mu}_k,\bm{\Sigma}_k\right) &=
  \frac{\exp\left[-\tfrac12(\bm{\mu}_k-\textbf{u}_m)^{\!\top}{\bm{\Sigma}_k}^{-1}(\bm{\mu}_k-\textbf{u}_m)\right]}
       {\sqrt{(2\pi)^4\det(\bm{\Sigma}_k)}},
\end{align}
where $\bm{\mu}_k = [I_k,x_k+{\Delta x}_k,y_k+{\Delta y}_k]$ is the mean representing the predicted intensity and coordinates of the MB and $\bm{\Sigma}_k=\mathrm{diag}(\sigma^2_{I,k},\sigma^2_{x,k},\sigma^2_{y,k})$ is the diagonal covariance matrix representing prediction uncertainties. The localisation loss is defined as the average negative log‑likelihood of each GT MB under a Gaussian mixture model (GMM) composed of per‑pixel predictive distributions $\mathcal{P}\left(\textbf{u}^{\text{GT}}_m | \bm{\mu}_k,\bm{\Sigma}_k\right)$. Each component is weighted by the normalised detection probability $\tilde{p}_k =p_k/{\sum^{K}_{j=1} p_j}$, ensuring the mixture weights sum to one across all pixels. The loss is formulated as:
\begin{align}
    \mathcal{L}_{\text{loc}} &=
  -\frac{1}{M}\sum_{m=1}^{M}
    \log\left[
      \sum_{k=1}^{K}
        \tilde{p}_k
        \mathcal{P}\left(\textbf{u}^{\text{GT}}_{m}|\bm{\mu}_k,\bm{\Sigma}_{k}\right)
    \right].
\end{align}

\begin{figure}[!t]
\centering
\includegraphics[width=1\linewidth]{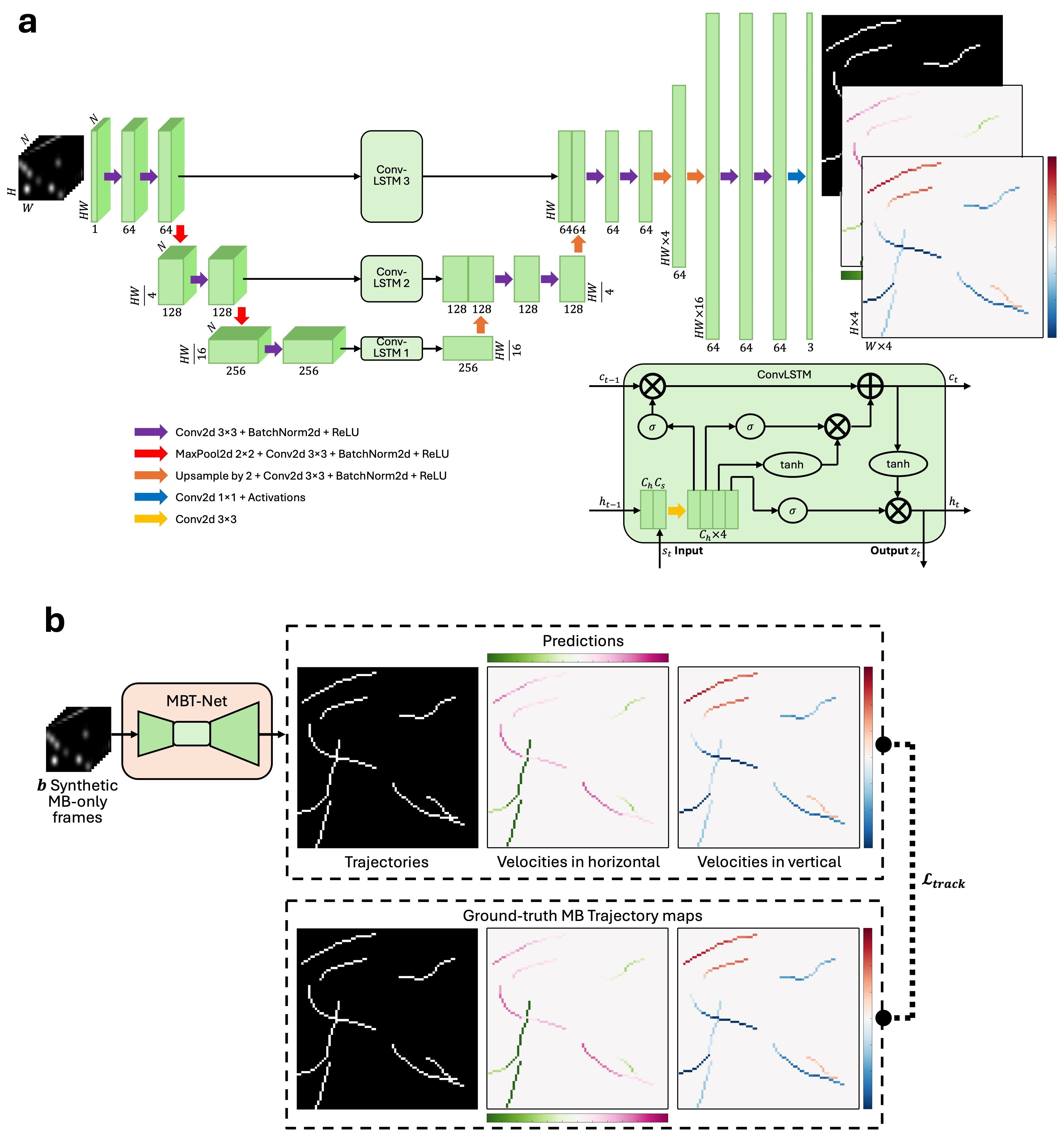}
%	\vspace{-0.3cm}
\caption{\textbf{Architecture and training losses of the MBT-Net.} \textbf{a}, Network architecture of the MBT-Net, highlighting the ConvLSTM block used to capture temporal dependencies across MB-only frames. \textbf{b}, Overall training strategy and corresponding loss functions employed for learning MB tracking in the MB-only domain.}
\label{fig:MBT_Net}
% \vspace{-0.3cm}
\end{figure}

\textbf{MBT-Net details.} To effectively capture both spatial and temporal information from the sequential CEUS input frames, our MBT-Net is designed based on a U-Net architecture combined with convolutional long short-term memory (LSTM) blocks inspired by \cite{chen2023localization}, as illustrated in Fig. \ref{fig:MBT_Net}. The encoder of the U-Net also includes two downsampling layers implemented via 2×2 max pooling, while the decoder features two upsampling layers achieved through bilinear interpolation followed by a 3×3 convolutional layer. LSTM is a type of recurrent neural network that uses gated cells to selectively retain or forget information over time, making it particularly effective at capturing and leveraging temporal dependencies in sequential data \citep{hochreiter1997long}. Convolutional LSTM (ConvLSTM), a variant of standard LSTM, replaces fully connected layers with convolutional operations, enabling better capture of spatiotemporal correlations, improved preservation of spatial structure, and reduced trainable parameters \citep{shi2015convolutional}. Exploiting this, three ConvLSTM blocks are integrated into the U-Net: two replace the standard skip connections and one substitutes the bottleneck module. To enhance the resolution of the resulting vasculature maps, the output from the U-Net is upsampled by a factor of 4 through two additional upsampling layers. As a result, MBT-Net takes a sequence of CEUS frames as input and outputs three channels: 1) a binary trajectory map $t$ indicating the presence of MB trajectories, 2) a velocity map $v_x$ and 3) a velocity map $v_y$ representing the horizontal and vertical components of MB velocities, respectively. 

To train the MBT-Net, we propose a composite loss function that integrates two components: a trajectory loss $\mathcal{L}_{\text{T}}$ and a velocity loss $\mathcal{L}_{\text{vol}}$: 
\begin{align}
    \mathcal{L}_{\text{MBT}} &= \mathcal{L}_{\text{T}} + \mathcal{L}_{\text{vel}}.
\end{align}
The trajectory loss $\mathcal{L}_{\text{T}}$ is used to supervise the detection of the presence of MB trajectories, formulated as a BCE loss between the predicted trajectory map $t$ and the GT one $t^{\text{GT}}$:
\begin{align}
    \mathcal{L}_{\text{T}}=
-\left[t^{\text{GT}}\log t + (1-t^{\text{GT}})\log(1-t)\right].
\end{align}
The velocity loss is designed to supervise the estimation of MB velocities along their trajectories based on the mean squared error (MSE) loss. However, due to the sparsity of the velocity maps, where valid velocity values are present only along the MB trajectories, the standard MSE becomes suboptimal. It will be dominated by zero regions and underweight the true values, which dilutes the gradients and weakens supervision in regions of interest. To address this, we propose a weighted MSE loss, where the weights are derived from the normalised detected trajectories, defined as $\tilde{t}_k = t_k / \sum^{K}_{i=1} t_i$. This weighting strategy emphasises areas where MB trajectories are detected, ensuring that the network focuses on accurate velocity prediction along these trajectories, thereby improving learning effectiveness and velocity estimation fidelity. As a result, the velocity loss $\mathcal{L}_{\text{vel}}$ is formulated as:
\begin{align}
    \mathcal{L}_{\text{vel}} = \sum_{k=1}^{K}
      \tilde{t}_k \left (
      \left\|
        v_{x,k} - v^{\text{GT}}_{x,k}
      \right\|_{2}^{2} + \left\|
        v_{y,k} - v^{\text{GT}}_{y,k}
      \right\|_{2}^{2} \right ).
\end{align}

\subsection{Training and inference}
Since the three proposed networks (MB-DT, MBL-Net, and MBT-Net) are fully convolutional with translation-equivariant filters, they can be trained on small image patches while being applied directly to full-size data at inference. Specifically, MB-DT was trained using patches of size 40×40 pixels randomly sampled from the target CEUS dataset. MBL-Net was trained on synthetic MB-only patches of the same size (40×40 pixels). Moreover, MBT-Net is trained on synthetic sequences of consecutive MB-only patches, with varied patch size  depending on the dataset: 40×40 pixels for the in vivo rabbit kidney dataset and 80×80 pixels for the ULTRA-SR Challenge dataset. The larger patch size was used for the ULTRA-SR dataset because the MBs exhibit larger displacements between frames, and a wider field of view is necessary to ensure that the MBs remain within the patch over time. Training networks on patches instead of full-size data not only reduces the memory usage, but also augments the effective sample size. More importantly, it allows for the selection of regions of interest (ROIs) in real datasets, enabling targeted training. For example, in the rabbit kidney dataset, we defined an ROI that outlines the kidney region and used only the patches within this area to train the MB-DT. This targeted approach enables the network to focus on relevant features while minimising the influence of irrelevant background, thereby improving both the effectiveness and robustness of the trained model.

\subsection{Synthetic data generation}
\textbf{MB generation.} To generate a synthetic MB-only patch, we follow four steps: 1) The number of MBs is randomly generated from a uniform distribution, $M \sim \mathcal{U}(1, M_{\max})$. 2) The centres of these MBs are uniformly randomly placed within the patch. 3) Each MB is assigned a random intensity $I_m \sim \mathcal{U}(I_{\min}, I_{\max})$. 4) Each MB is convolved with the predefined PSF. In this work, the PSF is estimated by fitting a 2D Gaussian to the average of 10 isolated MBs manually selected from the CEUS dataset.

\textbf{MB motion generation.} To simulate a sequence of consecutive MB-only patches, we extend the following steps: 1) Each MB in the first frame is assigned an initial velocity with a magnitude sampled from $V^{(1)}_{m} \sim \mathcal{U}(0, V_{\max})$ and a direction from $\theta^{(1)}_{m} \sim \mathcal{U}(-\pi, \pi)$. 2) For each subsequent frame, the MB positions are updated based on their current velocities using basic kinematics:
\begin{align}
    x^{(i+1)}_{m}=x^{(i)}_{m}+V^{(i)}_m\cos(\theta^{(i)}_m) \text{ and } y^{(i+1)}_{m}=y^{(i)}_{m}+V^{(i)}_m\sin(\theta^{(i)}_m).
\end{align}
3) The velocity of each MB is then perturbed by adding random variations to simulate natural motions:
\begin{align}
    V^{(i+1)}_m=V^{(i)}_m+\Delta V_{m} \text{ and } \theta^{(i+1)}_m=\theta^{(i)}_m+\Delta \theta_{m},
\end{align}
where the random variations are sampled from two Gaussian distributions: $\Delta V_{m} \sim \mathcal{N}(0, \sigma_{V})$ and $\Delta \theta_{m} \sim \mathcal{N}(0, \sigma_{\theta})$.
4) Repeat Steps 2 and 3 until the desired number of consecutive frames are generated.

\textbf{Implementation details.} The parameters mentioned above can be reasonably adapted according to the characteristics of the dataset. In this work, we empirically set $M_{\max}=10$ for patches of size 40×40 pixels and $M_{\max}=40$ for patches of size 80×80 pixels. The MB intensity range is defined as $I_{\min} = 0.05$ to exclude practically invisible MBs, and $I_{\max}=1$, as the dataset has been normalised to a maximum intensity of one. For the ULTRA-SR dataset, velocity parameters are set as $V_{\max}=78.28$ \si{\milli\meter/\second}, $\sigma_{V}=1.96$ \si{\milli\meter/\second} and $\sigma_{\theta}=\pi/18$. For the rabbit kidney dataset, the parameters are $V_{\max}=30.80$ \si{\milli\meter/\second}, $\sigma_{V}=3.08$ \si{\milli\meter/\second} and $\sigma_{\theta}=\pi/6$. It is worth noting that, due to the wide variability introduced by the random sampling process, the performance of the networks is robust to moderate changes of these parameters.

\subsection{Datasets for validation}
\textbf{In silico data.} The first dataset used for validation was the simulated data for the Ultrasound Localisation and Tracking Algorithms for Super Resolution (ULTRA-SR) Challenge \citep{lerendegui2022bubble}. The dataset comprises 500 synthetically generated frames that replicate acquisition with a high-frequency linear array L11-4 transducer operating at a 7.24 MHz centre frequency. Three-angle compounded plane-wave imaging was simulated with an angle step of 10°, resulting in a compounded frame rate of 50 Hz. Random noise added to the radio-frequency (RF) channel data was generated by filtering white Gaussian noise through the transducer bandwidth. Full simulation details are available in \cite{lerendegui2024ultra}.

\textbf{In vivo acquisition.} The in vivo datasets were acquired from a rabbit kidney using a GE HealthCare L3-12-D linear array (centre frequency 5 MHz; MI 0.1; single-cycle pulses) with ten-angle plane-wave compounding over -10° to 10°, yielding a compounded frame rate of 100 Hz. Full acquisition details are available in \cite{smith2025enhanced}. All procedures were carried out according to the Animals (Scientific Procedures) Act 1986 and received approval from the Animal Welfare and Ethical Review Body of Imperial College London.

\subsection{Evaluation metrics}
The performance of CycleULM is evaluated quantitatively using the following metrics. The contrast-to-noise ratio (CNR) is used to evaluate the contrast enhancement provided by the MB-DT. With $\bar{s}$ and $\bar{n}$ denoting the mean intensities in the selected signal and background regions, respectively, CNR (in dB) is expressed as:
\begin{align}
    CNR = 20\log_{10}\left(\frac{\bar{s}}{\bar{n}}\right).
\end{align}

With the GT available in the in silico dataset, MB localisation is evaluated using Recall, Precision and $F_1$ score, following \cite{lerendegui2022bubble,lerendegui2024ultra,yan2025enhancing}, given by:
\begin{align}
    &Recall=\frac{TP}{TP+FN}, \\
    &Precision=\frac{TP}{TP+FP}, \\
    &F_1=\frac{2TP}{2TP+FN+FP},
\end{align}
where $TP$ is the number of true-positive detected MBs, $FP$ represent the false-positive detections, and $FN$ represents the false negatives. Recall measures the proportion of actual MBs that are correctly localised; Precision measures the proportion of correctly localised MBs among all detected MBs; and F1 score provides a harmonic mean of Recall and Precision, offering a balanced evaluation of localisation performance. 

To further assess localisation accuracy, we introduce the mean localisation error, defined as the mean Euclidean distance between the positions of the detected MBs and the GT ones: 
\begin{align}
    \bar{E} = \frac{1}{M}\sum_{m=1}^{M}
\left\| \mathbf{P}_{d,m} - \mathbf{P}^{\mathrm{GT}}_{m} \right\|_2,
\end{align}
where $\mathbf{P}_{d,m}$ and $\mathbf{P}^{\mathrm{GT}}_{m}$ are the detected and GT coordinates of the $m$-th matched MB, respectively, and M is the number of matched (true-positive) pairs. 

Additionally, two image-based metrics are also computed between the target image $X$ and the GT image $Y$: root mean square error (RMSE) and the structural similarity index measure (SSIM) \citep{wang2004image}, given by:
\begin{align}
    &\mathrm{RMSE}(X,Y)
= \sqrt{\frac{1}{K}\sum_{k=1}^{K} \bigl(X_{k}-Y_{k}\bigr)^2}, \\ 
&\operatorname{SSIM}(X,Y)
= \frac{(2\mu_X \mu_Y + C_1)\,(2\sigma_{XY} + C_2)}
       {(\mu_X^2 + \mu_Y^2 + C_1)\,(\sigma_X^2 + \sigma_Y^2 + C_2)} ,
\end{align}
where $K$ is the number of pixels, $\mu_X$ and $\mu_Y$ are the means of the images, $\sigma_X^2$ and $\sigma_Y^2$ are their variances, $\sigma_{XY}$ is their covariance, and $C_1$ and $C_2$ are small stabilising constants. RMSE, the lower the better, measures the average magnitude of pixelwise intensity errors. SSIM, the higher the better, compares luminance, contrast, and structure between two images, ranging from 0 to 1.

\section*{Supplementary material}

\renewcommand{\thefigure}{S\arabic{figure}}
\setcounter{figure}{0}

\begin{figure}[!t]
\centering
% \vspace{-1.0cm}
\includegraphics[width=0.99\linewidth]{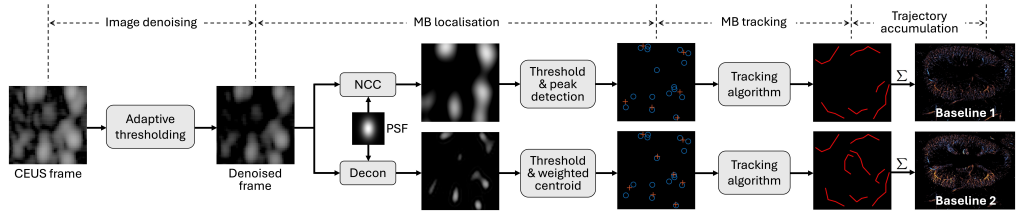}
% \vspace{-0.3cm}
\caption{\textbf{Traditional ULM post-processing pipeline.} }
\label{fig:S_1}
% \vspace{-0.3cm}
\end{figure}

Figure \ref{fig:S_1}

\begin{figure}[!t]
\centering
% \vspace{-1.0cm}
\includegraphics[width=0.99\linewidth]{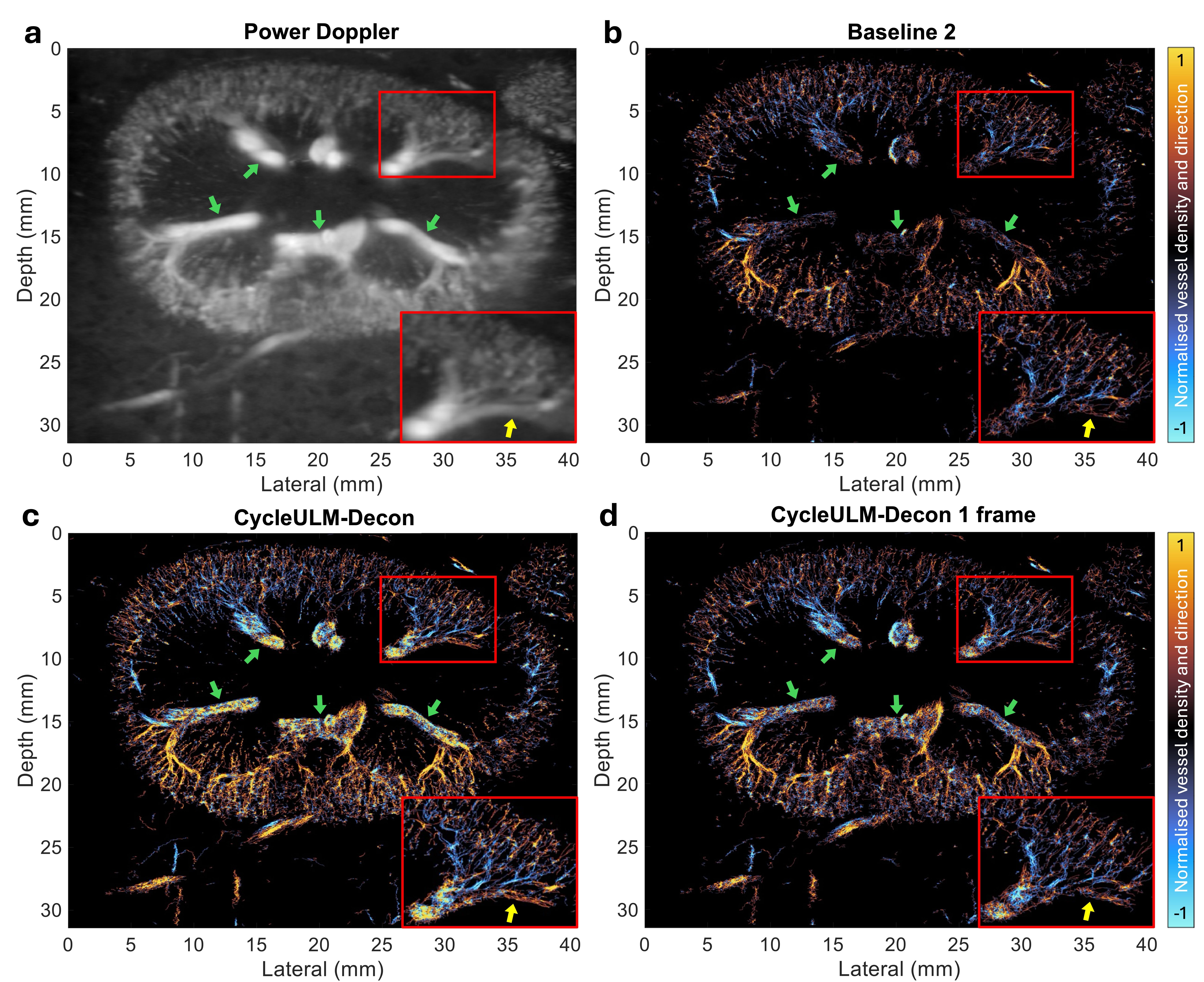}
% \vspace{-0.3cm}
\caption{\textbf{Evaluation of the 3-frame input design for MB-DT.} 
\textbf{a}, Power Doppler image created from 500 frames. 
\textbf{b}, SR map generated by the Baseline~2 method. 
\textbf{c,d}, SR maps generated by CycleULM-Decon when MB-DT is trained with three-frame input and single-frame input, respectively. 
Using three consecutive frames, CycleULM-Decon recovers finer vascular branches and achieves more complete vessel reconstruction.}

\label{fig:S_2}
% \vspace{-0.3cm}
\end{figure}

Figure \ref{fig:S_2}

Video 1 and Video 2: https://zenodo.org/records/18939887

%%Harvard
\bibliographystyle{elsarticle-harv.bst}\biboptions{authoryear}
\bibliography{refs}

\end{document}